\documentstyle[12pt,epsfig]{article}

\newcommand{\be}{\begin{eqnarray}}
\newcommand{\ee}{\end{eqnarray}}

\def\nue{{\nu_e}}
\def\anue{{\bar\nu_e}}
\def\numu{{\nu_{\mu}}}
\def\anumu{{\bar\nu_{\mu}}}
\def\nutau{{\nu_{\tau}}}

\def\lsim{\:\raisebox{-0.5ex}{$\stackrel{\textstyle<}{\sim}$}\:}

\newcommand{\ms}{\Delta m^2_{21}}
\newcommand{\ma}{\Delta m^2_{31}}

\newcommand{\sss}{\sin^2 \theta_{12}}
\newcommand{\sch}{\sin^2 \theta_{13}}

\newcommand{\sa}{\sin^2 \theta_{23}}

\newcommand{\mat}{\Delta m^2_{31}{\mbox {(true)}}}

\newcommand{\scht}{\sin^2 \theta_{13}{\mbox {(true)}}}

\newcommand{\sat}{\sin^2 \theta_{23}{\mbox {(true)}}}
\newcommand{\tmt}{$\theta_{23}$}
\newcommand{\tet}{$\theta_{13}$}
\newcommand{\tem}{$\theta_{12}$}

\newcommand{\pmm}{P_{\mu\mu}}
\newcommand{\sig}{$3\sigma$}

\def\gtap{\ \raisebox{-.4ex}{\rlap{$\sim$}} \raisebox{.4ex}{$>$}\ }

\begin{document}

\begin{flushright}
TIFR-TH/05-12\\
OUTP-0507P
\end{flushright}
\bigskip

\begin{center}
{\Large \bf Probing the deviation from maximal mixing of atmospheric 
neutrinos}

\vspace{.5in}

{\bf 
Sandhya Choubey$^{\star a}$ and Probir Roy$^{\dagger b}$}
\vskip .5cm
$^a${\it Rudolf Peierls Centre for Theoretical Physics,}
{\it University of Oxford,}\\
{\it 1 Keble Road, Oxford OX1 3NP, UK}\\
\vspace{.5cm}
$^b${\it Department of Theoretical Physics,} 
{\it Tata Institute of Fundamental Research, } \\
{\it Homi Bhaba Road, Mumbai 400 005, India }
\vskip 1in

\end{center}

\begin{abstract}
Pioneering atmospheric muon neutrino experiments have demonstrated the
near-maximal magnitude of the flavor mixing angle $\theta_{23}$.  But
the precise value of the deviation $D \equiv 1/2 - \sin^2 \theta_{23}$
from maximality (if nonzero) needs to be known, being of great
interest -- especially to builders of neutrino mass and mixing models.
We quantitatively investigate in a three generation framework the
feasibility of determining $D$ in a statistically significant manner
from studies of the atmospheric $\nu_\mu,\bar\nu_\mu$ survival
probability including both vacuum oscillations and matter
effects.  We show how this determination 
will be sharpened by considering the up-down
ratios of observed $\nu_\mu$- and $\bar\nu_\mu$-induced events 
and the differences of these ratios in specified
energy {\em and} zenith angle bins.  We consider 1 Megaton
year of exposure to a magnetized iron calorimeter such as the proposed
INO detector ICAL, taking into account both energy 
and zenith angle resolution functions.  
The sensitivity of such an exposure and the
dependence of the determination of $D$ on the
concerned oscillation parameters are discussed in detail.  
The vital use of matter effects in 
fixing the octant of $\theta_{23}$ is highlighted.
\end{abstract}

\vskip 1cm

\noindent $^\star$ email: sandhya@thphys.ox.ac.uk

\noindent $^\dagger$ email: probir@theory.tifr.res.in

\newpage

\section{Introduction}

Neutrino oscillation studies have come of age after the 
recent results from the 
Super-Kamiokande (SK) \cite{sklbye},
K2K \cite{k2k}, SNO \cite{sno2} and KamLAND \cite{kl2} 
experiments.
The existence of oscillations in atmospheric \cite{SKatm}
and accelerator 
generated \cite{k2k} 
$\numu$'s and $\anumu$'s with a modulus of squared mass 
difference\footnote{
We work in the conventional picture of three neutrino flavors with CPT 
conservation assumed. We also employ standard notation where 
$m_{ij}^2 = m_{i}^2 - m_{j}^2$ and $\theta_{ij}$ plus $\delta$ are 
defined by the PMNS matrix of our Eq. (\ref{upmns}) in \S 2.}
$|\ma| \sim 2.1 \times 10^{-3}$ eV$^2$ 
($1.4 \times 10^{-3}$ eV$^2 < |\ma| < 3.3 \times 10^{-3}$ eV$^2$ at 
\sig{}) and a mixing angle $\theta_{23} \sim 45^\circ$ 
($35^\circ < \theta_{23} < 54^\circ$ at \sig{}) as well as in 
solar \cite{sno2,solar} 
(reactor \cite{kl2}) $\nue$'s ($\anue$'s) with a squared mass 
difference$^1$ $\ms \sim 8\times 10^{-5}$ eV$^2$ 
($7.4\times 10^{-5}$ eV$^2 < \ms < 9.5 \times 10^{-5}$ eV$^2$ at  \sig{})
and a mixing angle $\theta_{12} \sim 34^\circ$ 
($30^\circ < \theta_{12} < 40^\circ$ at \sig{}) is now an 
accepted fact \cite{analyzes}. The goals of the next generation 
of experiments include both increased precision in the already 
measured oscillation parameters and the determination of the still 
unknown quantities such as $\theta_{13}$ (presently bounded by 
$0< \theta_{13} < 12^\circ$ at \sig{} \cite{analyzes,chooz}) 
as well as the sign 
of $\ma$ and eventually the CP violating phase $\delta$ in the 
PMNS matrix. An issue of great importance right now is the 
precise value of $\theta_{23}$ which dominantly controls 
$\numu$-$\nutau$ oscillations. The presently allowed range of $\theta_{23}$
does not enable one to distinguish it from its maximal\footnote{ 
Two flavor $\numu/\anumu$ oscillations in vacuum are controlled 
by $\theta_{23}$ only through the factor $\sin^22\theta_{23}$ 
which is highest when $\theta_{23}$ has the maximal value 
$\pi/4$. Matter effects and $\ms$-driven subdominant 
atmospheric neutrino oscillations have different dependence on 
$\theta_{23}$.} value $\pi/4$.
A naturally arising question is: how can one significantly narrow
down this range in the future and possibly detect a nonzero deviation
\be
D \equiv \frac{1}{2} - \sa
\label{Eq:D}
\ee
from maximal mixing. This is what we consider in the context of 
forthcoming\footnote{An analysis of the allowed range of $D$, based on 
extant atmospheric $\numu+\anumu$ data, is available in 
\cite{maltonimax23}.}
studies of atmospheric $\numu$'s and $\anumu$'s.

Let us mention some current theoretical ideas regarding $\theta_{23}$.
A simple way of understanding a maximal value for $\theta_{23}$ 
is to invoke the concept of a $\mu$-$\tau$ 
exchange symmetry \cite{mohapatra-werner}. Though this 
immediately leads to $\theta_{23}=\pi/4$, it also yields
at the same time a vanishing \tet{} and CP conservation 
in the neutrino sector. While the last two features 
could in principle turn out 
to be the facts of Nature, our current hopes are to the 
contrary. A nonzero and measurable \tet{} as well as the occurrence 
of CP violation in the neutrino sector can lead to much that is 
interestingly new in neutrino physics and can be studied in 
laboratory experiments with reactor, accelerator, atmospheric,
solar and supernova neutrinos. The general expectation 
\cite{mohapatra-werner}, therefore, is that any $\mu$-$\tau$ 
interchange symmetry, if present, must be broken so as to 
allow a nonvanishing \tet{} and (possibly observable) CP violation
in the neutrino sector. Such a breaking would generally cause 
$D$ to be different from zero. However, its magnitude and sign 
would be given by the yet undetermined symmetry breaking mechanism.
Another idea has been that of quark-lepton complementarity \cite{qlc}.
In this approach a bimaximal neutrino Majorana mass matrix 
is ``rotated'' by a unitary matrix diagonalizing charged 
left chiral leptons in generation space. The latter is postulated 
to be the same as the CKM matrix resulting in a reduction of \tem{} 
from $\pi/4$ by an amount comparable to the Cabibbo angle 
$\theta_C \sim 13^\circ$ in conformity with observation. In this 
case \tmt{} also is expected to be reduced from $45^\circ$ by an 
amount of the order of $\sin^{-1}s_{23}^{\rm CKM} \sim 2.4^\circ$
so that $D$ is expected to be positive and in the vicinity of 0.04. 
However, variants \cite{qlc} of this hypothesis exist with D
predicted to be significantly different from 0.04.
There are many other model predictions, utilizing flavor symmetries,
GUT relations and textures, covering $|D|$ from 0.005 to $\geq 0.16$,
as listed in Table II of \cite{antusch}.

We thus see that an experimental measurement of $D$ would be of 
great interest to the builders of neutrino mass matrix models.
Apart from the magnitude of $|D|$, $sgn(D)$ is also of 
importance. Its determination would fix the octant of \tmt, 
{\it i.e.} whether \tmt{} exceeds or is less than $\pi/4$. With the 
goal of determining $D$, 
one can explore different experimental options.
A detailed survey of the 
capabilities of forthcoming and futuristic 
accelerator based long baseline neutrino/antineutrino experiments
along this line has already been given in \cite{antusch,mina23}.
Here we want to consider future atmospheric neutrino studies 
guided by our knowledge of extant ones \cite{maltonimax23}.
A major new venture in this direction will be the proposed 
Hyper-Kamiokande/UNO/MEMPHYS type of experiments \cite{megaton} with 
Megaton water Cerenkov detectors. Detailed studies 
\cite{hubert2k+hk,mega1,mega2}
have been made (among other things) of the $\numu+\anumu$ 
survival probability, information on which will emanate from 
such a detector. While this kind of data can yield information 
also on $D$ \cite{maltonimax23}, 
it will be largely from the vacuum oscillations of 
$\numu$ and $\anumu$ and will be less sensitive to matter effects.
Depending on the mass ordering of the neutrinos (normal or inverted), 
larger matter effects appear in the neutrino or antineutrino channel 
for long baseline distances ($L > 1000$ km). 
In addition, the sign of the earth matter
effect term in the difference of survival probabilities 
$\Delta P_{\mu\mu}$ (as defined in $\S$2) 
depends on the energy and zenith angle of the atmospheric
$\nu_\mu/\bar\nu_\mu$.
Since water Cerenkov atmospheric neutrino experiments
measure the $\numu+\anumu$ survival probability and since their 
poor detector resolution means that the 
data collected would have to 
be grouped into very broad sub-GeV and multi-GeV 
energy bins as in the SK data sample, matter effects
in this class of experiments will be partially washed out.
In order to maximize the extraction of information from the 
individual survival probabilities of atmospheric $\numu$'s and 
$\anumu$'s, propagating both through vacuum and through earth,
one needs to measure them separately as a function of energy and 
baseline length. 
In fact, 
the individual survival probabilities will be measurable in 
a large magnetized iron calorimeter such as the ICAL detector 
of the proposed INO experiment \cite{ino}. 
This experiment is expected to have a good 
detector resolution allowing fine binning of the data in 
energy and zenith angle, essential for gleaning out the 
$D$ dependence of earth matter effects.
Our aim is to focus on 
the feasibility of utilizing the data in future from such a 
detector to determine $D$. 

There have already been several studies \cite{sergio,indu,pomita}
of the possibility of determining the normal or inverted nature 
of the mass ordering of neutrinos, {\it i.e.} $sgn(\ma)$, in 
magnetized iron calorimetric detectors. The question of the 
determination of $D$ has been touched upon \cite{indu} 
but not analyzed in detail. In this paper we carry out a 
detailed study of the feasibility of a measurement of both 
the magnitude and the sign of $D$ from data simulated 
for 1 Megaton year of exposure of a large magnetized iron calorimeter,
such as ICAL, to atmospheric muon neutrinos and antineutrinos.
In principle, a magnetized iron calorimetric detector in an underground
observatory (such as the proposed INO laboratory) could also study other
types of events: upward going muons and electron neutrino events. As a
first step, we would like to restrict 
our studies to fully contained events
only and hence do not consider upward going muons,
which are much harder to incorporate. 
Furthermore, the large
thickness \cite{ino}
of the iron plates of the proposed ICAL detector precludes
the detection of electrons and will not trigger on to events induced by
$\nu_e$s and $\bar \nu_e$s. This is why we confine ourselves to
atmospheric $\nu_\mu, \bar \nu_\mu$ studies. Furthermore, ICAL is designed
to have a detection threshold of neutrino energy = 1 GeV and therefore
cannot study sub-GeV events. This is why we
consider the multi-GeV regime only.
We take a three generation system and use exact numerical 
solutions of the neutrino (and antineutrino) equations of 
motion in the atmosphere and in earth matter using the 
PREM \cite{PREM} density profile as an input. We assume what we 
regard as reasonable energy and zenith angle resolution 
functions for the detector and concentrate on up-down 
asymmetry ratios \cite{indu} $U_N/D_N$ and $U_A/D_A$ for 
$\numu$- and $\anumu$-induced events in appropriate 
energy and zenith angle bins. We are then able to pinpoint 
the precise bins from where the utilizable information on 
{\it both} vacuum oscillations and matter effects can be 
extracted -- leading to a determination of $D$. We also show 
how the use of matter effects through the difference 
$U_N/D_N - U_A/D_A$ will enable one to fix the sign of $D$ and 
resolve the octant ambiguity in $\theta_{23}$.

The paper is organized in the following way. Section 2
contains a slightly simplified and approximate analytical 
treatment of the survival probabilities of $\numu$'s and 
$\anumu$'s propagating in matter of constant density;
this is to bring out the basic physics issues in question, 
specifically, highlighting the phenomenon of resonance.
In \S 3 we present our numerical results on the $\numu$ 
survival probability (after propagation in earth matter)
as it varies with the energy $E$ and baseline length $L$; 
in particular, the occurrence of extrema in this variation 
is highlighted through the definitions of SPMIN1, SPMAX and 
SPMIN2 which appear. Section 4 comprises the methodology 
for the extraction of the up-down asymmetry ratios 
$U_N/D_N$ and $U_A/D_A$ from the simulated data. The details of 
our statistical ($\chi^2$) analysis are presented in \S 5.
The determination of 
$|D|$ from the simulated data and the 
estimated intervals on both sides of 
$D=0$ for which this will be possible at the \sig{} level are given 
in \S 6. The procedure for the fixation of the sign of $D$
(and hence the octant of $\theta_{23}$) is described in \S 7.
In \S 8, which contains our summary and conclusions, we also
provide a discussion of the sensitivity of an 
ICAL-like detector to $D$, 
as compared to the forthcoming long baseline and water 
Cerenkov detector studies with accelerator generated and 
atmospheric neutrinos respectively.

\section{\label{sec:probanalytical}
Approximate analytical treatment of $\nu_\mu$/$\bar\nu_\mu$ survival
probability in matter}

In the standard parametrization, the Pontecorvo-Maki-Nakagawa-Sakata 
\cite{PMNS} matrix is given by
\be 
U = \pmatrix
{c_{12} c_{13} & s_{12} c_{13} & s_{13} e^{-i \delta} 
\cr
-s_{12} c_{23} - c_{12} s_{23} s_{13} e^{i \delta} 
& c_{12} c_{23} - s_{12} s_{23} s_{13} e^{i \delta} 
& s_{23} c_{13} \cr
s_{12} s_{23} - c_{12} c_{23} s_{13} e^{i \delta} 
& 
c_{12} s_{23} - s_{12} s_{23} s_{13} e^{i \delta} 
& c_{23} c_{13}\cr
} 
.
\label{upmns} 
\ee
In Eq. (\ref{upmns}) $c_{ij} = \cos\theta_{ij}$, $s_{ij} =
\sin\theta_{ij}$ and, while the phase $\delta$ has been retained,
the two Majorana phases have been ignored since they do not contribute
to neutrino oscillations.  The effective Hamiltonian for such
oscillations in matter with varying density can be expressed in the
flavor basis as a function of the path length $x$ of the neutrino from
its source.  Thus 
\be
H (x) = {1 \over 2E} U \pmatrix{0 & 0 & 0 \cr 0 & \ms & 0 \cr 0 &
0 & \ma} 
U^\dagger + \pmatrix{V(x) & 0 & 0 \cr 0 & 0 & 0 \cr 0 & 0 & 0} .
\label{hammatt}
\ee
In Eq. (\ref{hammatt}), 
$\Delta m^2_{ij} \equiv m^2_i - m^2_j$, $m_i$ and $E$ being
the mass of the $i$th (physical) neutrino and the neutrino energy
respectively, whereas the potential $V(x)$ is given by
\be
V(x) = \sqrt{2} G_F N_e (x).
\label{three}
\ee
Here $N_e(x)$ is the electron density of matter, so that the potential
can be rewritten as
\be
V(x) = (7.56 \times 10^{-14}) {\rho(x) \over {\rm gms/cc}} 
Y_e (x)~ {\rm eV},
\label{four}
\ee
$\rho(x)$ being the density of earth matter in the path of the
neutrino and $Y_e (x) \ (\simeq 0.5)$ being the number of electrons
per nucleon in the same.

The neutrino evolution operator $S(x,0)$ has the matrix element
\be
S_{\gamma\beta} (x,0) = \langle \nu_\gamma(x) |\nu_\beta(0)\rangle ,
\label{five}
\ee
$\beta,\gamma$ being flavor indices.  This obeys the evolution
Equation
\be
i {dS_{\gamma\beta} (x,0) \over dx} = [H(x),S(x,0)]_{\gamma\beta}.
\label{six}
\ee
The probability for a neutrino flavor transition $\beta \rightarrow
\gamma$ at a baseline length $L$ is given by 
\be
P_{\beta\gamma} (L) \equiv P[\nu_\beta (0) \rightarrow \nu_\gamma (L)]
= |S_{\gamma\beta} (L,0)|^2.
\label{seven}
\ee
In case the earth matter density $\rho$ is taken\footnote{This is an
inaccurate assumption for the passage of GeV and multi-GeV
atmospheric neutrinos through the earth when $L$ exceeds 1000 km.
However, we need this assumption only for displaying analytic expressions. Our
numerical results are obtained without assuming a constant $\rho$, and 
actually, with the PREM \cite{PREM} 
earth matter density profile as an input.
Moreover, they are correct to all orders in $\alpha \equiv 
\ms/\ma$ and $s_{13}$.} to be constant between the
production and detection points of the neutrino, the potential $V$
also becomes a constant.  It is then useful to diagonalize the
Hamiltonian of Eq. (\ref{hammatt}) with eigenvalues $\lambda_{1,2,3}
(2E)^{-1}$. 
\be
H = {1 \over 2E} U^M \pmatrix{\lambda_1 & 0 & 0 \cr 0 & \lambda_2 & 0
\cr 0 & 0 & \lambda_3} U^{M^\dagger},
\label{eight}
\ee
$U^M$ being the lepton mixing matrix in matter.  Then one can write 
\be
S_{\gamma\beta} (L,0) &=& \sum^3_{i=1} U^{M^\star}_{\gamma i}
e^{-i\lambda_i L(2E)^{-1}} U^M_{\beta i},
\\
P_{\beta\gamma} (L) =& \delta_{\beta\gamma}& - 4 \sum_{j>1} \Re
\left(U^M_{\beta i} U^{M^\star}_{\gamma i} U^{M^\star}_{\beta j}
U^M_{\gamma j}\right) \sin^2 {\Delta m^2_{ij} L \over 4E} 
\nonumber \\
&&+ 2 \sum_{j>1} \Im \left(U^M_{\beta i} U^{M^\star}_{\gamma i}
U^{M^\star}_{\beta j} U^M_{\gamma j}\right) \sin {\Delta m_{ij}^2 L
\over 2E}.
\ee

\begin{table}[p]
\begin{center}
\begin{tabular}{|c|c|c|c|}
\hline
$E$ in GeV & $L$ in km & $A$ & $A\Delta$ \\
\hline
1 & 1000 & 0.099 & 0.252 \\
1 & 3000 & 0.125 & 0.953 \\
1 & 5000 & 0.131 & 1.673 \\
1 & 7000 & 0.155 & 2.754 \\
1 & 9000 & 0.171 & 3.908 \\
1 & 11000 & 0.229 & 6.409 \\
\hline
3 & 1000 & 0.297 & 0.252 \\
3 & 3000 & 0.375 & 0.953 \\
3 & 5000 & 0.395 & 1.673 \\
3 & 7000 & 0.465 & 2.754 \\
3 & 9000 & 0.513 & 3.908 \\
3 & 11000 & 0.688 & 6.409 \\
\hline
5 & 1000 & 0.495 & 0.252 \\
5 & 3000 & 0.626 & 0.953 \\
5 & 5000 & 0.659 & 1.673 \\
5 & 7000 & 0.775 & 2.754 \\
5 & 9000 & 0.855 & 3.908 \\
5 & 11000 & 1.147 & 6.409 \\
\hline
7 & 1000 & 0.693 & 0.252 \\
7 & 3000 & 0.876 & 0.953 \\
7 & 5000 & 0.922 & 1.673 \\
7 & 7000 & 1.084 & 2.754 \\
7 & 9000 & 1.197 & 3.908 \\
7 & 11000 & 1.606 & 6.409 \\
\hline
9 & 1000 & 0.892 & 0.252 \\
9 & 3000 & 1.126 & 0.953 \\
9 & 5000 & 1.185 & 1.673 \\
9 & 7000 & 1.394 & 2.754 \\
9 & 9000 & 1.538 & 3.908 \\
9 & 11000 & 2.064 & 6.409 \\
\hline
\end{tabular}
\caption{\label{tab:matter}
Variation of $\Delta$ and $A\Delta$ with $E$ and $L$ assuming
$\rho$ to be constant and equal to 4.52 gms/cc.}
\end{center}
\end{table}
More specifically, the muon neutrino and antineutrino survival
probabilities are given by 
\be
P_{\mu\mu} (L) = 1 - 4 \Bigg(|U^M_{\mu 1}|^2 |U^M_{\mu 2}|^2 \sin^2
{\lambda_1 - \lambda_2 \over 4E} L &+& |U^M_{\mu 1}|^2 |U^M_{\mu 3}|^2 \sin^2
{\lambda_1 - \lambda_3 \over 4E} L 
\nonumber \\
 \nonumber \\
&+& |U^M_{\mu 2}|^2 |U^M_{\mu 3}|^2 \sin^2 {\lambda_2 - \lambda_3 \over
4E} L\Bigg),
\ee
\be
P_{\bar\mu \bar\mu} (L) = P_{\mu\mu} (L,V \rightarrow -V).
\ee

\noindent
The elements of $U^M$, appearing in Eqs. (9), can be parametrized in
the same way as the `vacuum' mixing matrix $U$ of Eq. (\ref{upmns}) 
except that
one needs to use the corresponding values of the angles and the phase
in matter, i.e. $\theta_{ij} \rightarrow \theta^M_{ij}$ and $\delta
\rightarrow \delta^M$.

Tractable analytic expressions $P_{\alpha\beta} (L)$ emerge only after
some additional approximations.  For the purpose of displaying the
relevant analytic expressions, in addition to assuming the constancy$^4$
of $\rho$ and hence of $V$, we choose to neglect $O (s^3_{13})$ and
$O(\alpha^2)$ terms, $\alpha$ being $\ms/\ma$. 
From what was discussed in \S 1, we already know that
\be
s^3_{13} < 0.008, 
\\
\alpha^2 \simeq 0.001.
\ee
Therefore these approximations do not generally make any significant
practical difference from the exact numerical results, as will be seen
later. On the other hand, they do enable us to display the two
quantities of our interest, namely $P^{vac}_{\mu\mu}$, which is the
muon neutrino survival probability in vacuum, and $\Delta
P_{\mu\mu} (L) \equiv P_{\mu\mu} (L) - P_{\bar\mu \bar\mu} (L)$, which
is the difference between the muon neutrino and antineutrino survival
probabilities in matter, in
analytic form, making their physical features rather transparent.  The
eigenvalues of $\lambda_{1,2,3}$ and the elements of the matrix $U^M$
of Eq. (\ref{eight}) can be calculated within the above mentioned
approximations following the method described in Ref. \cite{[R2]}. They
can then be substituted in Eq. (11) to compute $P^{vac}_{\mu\mu} (L)$
and $\Delta P_{\mu\mu} (L)$ for atmospheric neutrinos.   

In order to display the desired analytic expressions, we first define
a dimensionless quantity 
\be
\Delta \equiv {\ma L \over 4E}.
\ee
The expression for $P^{vac}_{\mu\mu}$ can now be given as \cite{[R2]}
\be
P^{vac}_{\mu\mu} = 1 \!\!\!&-&\!\!\! 4s^2_{23} c^2_{23}(1 - c^2_{13} s^2_{23}) 
\sin^2\Delta \nonumber \\[2mm]
\!\!\!&+&\!\!\!  4\alpha c_{12} c_{23} (c_{12} c_{23} -
2s_{13}s_{12}s^3_{23}\cos\delta) \Delta \sin 2\Delta + {\cal
O}(\alpha^2,s^3_{13}).    
\ee
For the description of matter effects, it is convenient to define
another dimensionless quantity
\be
A \equiv {2E V \over \ma} .
\ee 
For atmospheric muon neutrinos of GeV and multi-GeV energies and
low to large pathlengths, the magnitude of $A$ varies from about 0.1
to about 2.1.  On the other hand, the energy independent $A\Delta$
varies from about 0.25 to about 6.4 for $\rho \simeq 4.52$ gm/cc,
cf. Table \ref{tab:matter}, 
where the values of these parameters have been given for
various choices of $E$ and $L$.  The expression for $\Delta
P_{\mu\mu} (L)$ now reads \cite{[R2]}
\be
\Delta P_{\mu\mu} (L) &=& 4s^2_{13} s^2_{23}
\left[{\sin^2(1+A)\Delta \over (1+A)^2} - {\sin^2(1-A)\Delta \over
(1-A)^2}\right] \nonumber \\[2mm] 
&& -8s^2_{13} c^2_{23} s^2_{23} \left[\sin\Delta \cos
A\Delta \left\{{\sin(1+A)\Delta \over (1+A)^2} - {\sin(1-A)\Delta
\over (1-A)^2}\right\} + \Delta \sin 2\Delta {A \over 1-A^2} \right]
\nonumber \\[2mm] 
&& -8\alpha s_{13} c_{12} s_{12} c_{23} s_{23}
\cos\delta \Bigg[\cos\Delta {\sin A\Delta \over A}
\left\{{\sin(1+A)\Delta \over 1+A} - {\sin(1-A)\Delta \over
1-A}\right\} \nonumber \\[2mm]
&& + (c^2_{23} - s^2_{23}) \sin\Delta \left\{\sin\Delta
{2A \over 1-A^2} + {\sin A\Delta \over A} \left({\cos(1+A)\Delta \over
1+A} - {\cos(1-A)\Delta \over 1-A}\right)\right\}\Bigg] 
\nonumber \\
&&+ O(\alpha^2,s^3_{13}). \hspace*{9cm}
\label{Eq:deltapmm}
\ee
This expression vanishes for $s_{13} = 0$, clearly showing the need
for the latter to have a nonzero value in order to make $\Delta
P_{\mu\mu}$ nonzero.    
If the limit $\alpha \rightarrow 0$ is taken and the ``small $A$''
approximation is used\footnote{From Table \ref{tab:matter}, the linear $A$
approximation would appear to be accurate to within 90\% only for
neutrino energies 
below 3 GeV {\it and} pathlengths below 9000 km.}, retaining only
linear terms in $A$, Eq. (\ref{Eq:deltapmm}) reduces to
\be
\Delta P_{\mu\mu} \simeq -16A(1/2 - |U_{\mu 3}|^2) |U_{e3}|^2
|U_{\mu3}|^2 (\Delta \sin 2\Delta - 2 \sin^2 \Delta)
\label{Eq:deltapmm2}
\ee
in agreement with the result first reported in Ref. \cite{[R3]}.  In this
approximate limit, though not in general, $\Delta P^A_{\mu\mu}$ is
directly proportional to $|U_{e3}|^2 (1/2 - |U_{\mu 3}|^2)$.  The more
general dependence on $s_{13}$ and $D$ is, of course, hidden in 
Eq. (\ref{Eq:deltapmm}).

The last-mentioned $\alpha \rightarrow 0$ limit may in fact be more
relevant than being a mere device to get to Eq. (\ref{Eq:deltapmm2}).  
Given (1) the
accidental feature that the atmospheric $\nu_\mu/\nu_e$ ratio would
have been $\sim 2$ in the relevant energy range if $\nu_\mu
\rightleftharpoons \nu_\tau$ oscillations were absent and (2) the fact of a
near-maximal $\theta_{23}$, the effect of the solar neutrino squared
mass difference $\ms$ on
atmospheric neutrino oscillations can be shown to be small \cite{[R4]}.  
It may therefore be useful to display also the analytical expression for
the survival probabilities of muon neutrinos and antineutrinos with
the assumptions of only the constancy of $\rho$ and the neglect of
${\cal O}(\alpha)$ terms.  Taking the limit $\alpha
\rightarrow 0$ and using the notation of
Eqs. (16) and (18), one then has
\be
\lim_{\small\alpha~\rightarrow~0} \lambda_1 &=& {1\over2}
\left[\ma (A+1) - {(\ma)}^M\right], 
\nonumber \\
\lim_{\small\alpha~\rightarrow~0} \lambda_2 &=& 0, 
\nonumber \\
\lim_{\small\alpha~\rightarrow~0} \lambda_3 &=& {1\over2}
\left[\ma (A+1) + {(\ma)}^M\right], 
\label{eq:lambda}
\ee
for neutrinos with
\be
{(\ma)}^M = \ma |(A-1)| \left[1 + 4s^2_{13}
(A-1)^{-2}\right]^{1/2} 
\ee
and replacing $A$ by $-A$ for antineutrinos.  Furthermore, it follows
that
\be 
\lim_{\small\alpha~\rightarrow~0} |U^M_{\mu 1}| &=& |s_{23} s_{13}^M|,
\\
\lim_{\small\alpha~\rightarrow~0} |U^M_{\mu 2}| &=& |c_{23}|,
\\
\lim_{\small\alpha~\rightarrow~0} |U^M_{\mu 3}| &=& |s_{12} c^M_{13}|,
\ee
where $s^M_{13} \equiv \sin\theta^M_{13}$, $c^M_{13} =
\cos\theta^M_{13}$ and $\theta^M_{13}$ is given in terms of
$\theta_{13}$ and $A$ by the relation
\be
\sin^2 2\theta^M_{13} = \sin^2 2\theta_{13} \left({\ma
\over {(\ma)}^M}\right)^2 = {\sin^2 2\theta_{13} \over (A -
\cos 2\theta_{13})^2 + \sin^2 2\theta_{13}}.
\label{eq:sinm}
\ee
Eqs. (20) -- (25) enable us to rewrite Eqs. (12) and (13) in the limit $\alpha
\rightarrow 0$ as 
\be
\lim_{\small\alpha~\rightarrow~0} P_{\mu\mu} (L) = 1 - 
P^1_{\mu\mu} (L) - P^2_{\mu\mu} (L) -
P^3_{\mu\mu} (L), 
\label{eq:p22constrho}
\ee
\be
P_{\bar\mu \bar\mu} (L) = P_{\mu\mu} (L,A \rightarrow -A),
\ee
with
\be
P^1_{\mu\mu} (L) = \sin^2 \theta^M_{13} \sin^2 2\theta_{23} \sin^2
{\ma (A+1) - {(\ma)}^M \over 8E} L,
\label{eq:p22term1}
\ee
\be
P^2_{\mu\mu} (L) = \cos^2 \theta^M_{13} \sin^2 2\theta_{23} \sin^2
{\ma (A+1) + {(\ma)}^M \over 8E} L,
\label{eq:p22term2}
\ee
\be
P^3_{\mu\mu} (L) = \sin^2 2\theta^M_{13} \sin^4 \theta_{23} \sin^2
{{(\ma)}^M \over 4E} L.
\label{eq:p22term3}
\ee
We note that $P^1_{\mu\mu} (L)$ vanishes in vacuum (when matter
effects go to zero), but $P^{2,3}_{\mu\mu}$ do not.  
In \S 3, we shall numerically study the behavior of $P^{1,2,3}_{\mu\mu}
(L)$ as functions of the neutrino energy $E$ for different baseline
lengths $L$.  The expression for $\sin^2 2\theta^M_{13}$ in Eq. 
(\ref{eq:sinm})
shows the effect of the MSW resonance when $A$ equals $\cos
2\theta_{13}$, i.e.
\be
E = E_{\rm res.} \equiv {\ma \cos 2\theta_{13} \over
2\sqrt{2} G_F N_e}.
\label{eq:eres}
\ee
Though $E_{\rm res.}$ of Eq. (\ref{eq:eres}) 
has no explicit dependence on $L$,
in practice an implicit $L$ dependence creeps in through $N_e$ for
path lengths involving significant spatial variations of the earth's
density.  Furthermore, at or near the resonance, Eq. (\ref{Eq:deltapmm}) 
would not be
trustworthy since, with $s_{13}$ small, $A$ is rather close to
unity -- making some of the RHS terms blow up.
At $E = E_{\rm res.}$, $\sin^2 2\theta^M_{13}$ reaches its
maximum value, i.e. unity.  For $\ma = (2-3) \times 10^{-3} \ {\rm 
eV}^2$ and a small 
$\theta_{13}$, the resonance could occur \cite{[R5]} for atmospheric
muon neutrinos passing through the earth with energies between 5 and
10 GeV.  At resonance, the conversion probability of a muon neutrino
becomes sizable even for a small $\theta_{13}$.  The effect of a
small nonzero $\theta_{13}$ could then be observed as an excess of
upward going electron neutrinos.

\section{\label{sec:probnumerical}
Numerical results on $\numu$ survival probability in matter}

\begin{figure}[t]
\begin{center}
\includegraphics[width=16.0cm, height=12cm]
{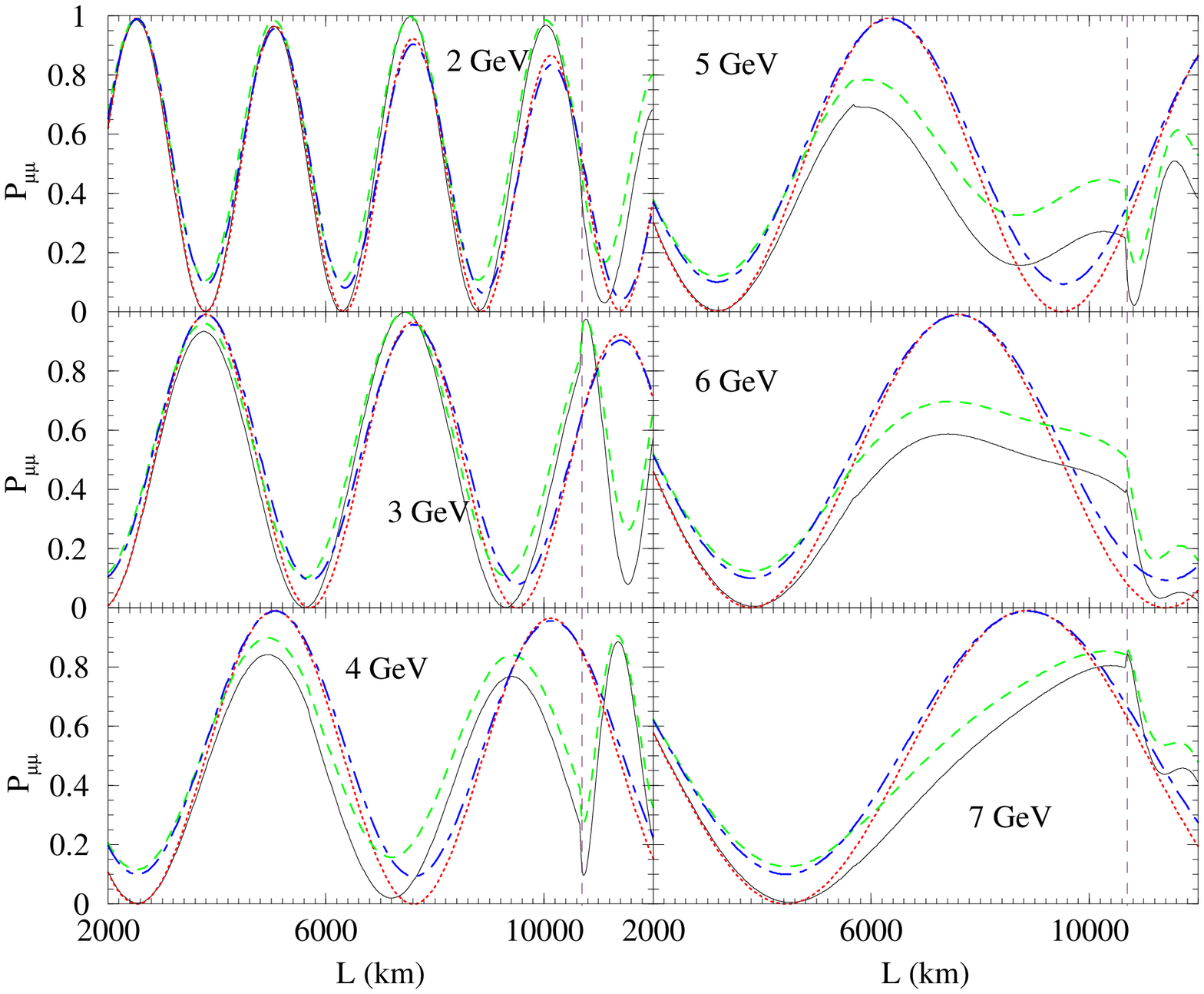}
\caption{$P_{\mu\mu}$ as a function of $L$ with 
fixed $E$ (and at various values of $E$) for neutrinos travelling in
vacuum and matter.  The black 
solid lines and the dashed green lines are for propagation in matter
with $\sin^2 
\theta_{23} = 0.5$ and 0.36 respectively.  The red dotted lines and
the blue dot-dashed lines are for that in vacuum for the same
respective values of $\sin^2 \theta_{23}$. 
The vertical dashed lines 
represent the mantle-core boundary inside the earth.}
\label{fig:p22fixedE}
\end{center}
\end{figure}

We now present our results for the muon 
neutrino survival probability $\pmm$ in vacuum and in matter.  To
obtain these, we have numerically solved 
the three-generation differential equation of motion of
the neutrinos exactly, using the PREM profile for the earth 
matter densities \cite{PREM}. 
Fig. \ref{fig:p22fixedE} shows $\pmm$ as a function of
$L$, for six different fixed values of $E$.  This plot (as well as
subsequent plots presented in this paper except where specified
otherwise) has been generated with chosen benchmark values of the
other concerned oscillation parameters $\Delta m^2_{31}$, $\Delta
m^2_{21}$, $\sin^2\theta_{12}$, $\sin^2\theta_{13}$ and $\delta$ -- as
tabulated in Table \ref{tab:benchmark} -- and two values of $\sa$ 
as explained in the caption. The following 
features are evident from Fig. \ref{fig:p22fixedE}.

\begin{table}[h]
\begin{center}
\begin{tabular}{|l|}
\hline
\\[.5mm]
$\Delta m^2_{31} = 2 \times 10^{-3} \ {\rm eV}^2$ \\[2mm]
$\Delta m^2_{21} = 8 \times 10^{-5} \ {\rm eV}^2$ \\[2mm]
$\sin^2\theta_{12} = 0.28$ \\[2mm]
$\sin^22\theta_{13} = 0.1$ \\[2mm]
$\delta = 0$ \\
\hline
\end{tabular}
\caption{\label{tab:benchmark}
Chosen benchmark values of oscillation parameters, except
$\theta_{23}$, with assumed normal mass ordering.}
\end{center}
\end{table}

\begin{itemize}
\item 
Matter effects inside the mantle begin to be significant for  
$E \gtap 4$ GeV,  $L \gtap 4000$ km 
and are largest for $E$ between 5 and 6 
GeV, reducing a bit for $E$ = 7 GeV.
\item For $E \gtap 4$ GeV and inside the mantle, the survival probability
in matter at the peaks invariably reduces from its value of near unity
in vacuum. This effect increases as $\sa$ is increased 
from 0.36 to 0.5 and beyond.
\item For $E \gtap 4$ GeV and inside the mantle, matter effects tend 
to increase 
the survival probability at the troughs from its vacuum
value of $P_{\mu\mu} = 1 - 4 c_{13}^2 c_{23}^2 (1 - c_{13}^2 c_{23}^2)$.
This increase however reduces as the value of $\sa$ is raised from
0.36 to 0.5 and beyond.
\item
For $E \gtap 4$ GeV and  $L$  large but within the 
mantle (specifically between 8000 km and 10000 km), the matter
contribution to the survival probability, i.e. 
$P_{\mu\mu} - P_{\mu\mu}^{vac}$, can change sign.
\item 
The survival probability in matter has a fairly 
sharp discontinuity at the mantle-core boundary and behaves very 
differently with distance inside the core as compared with the 
mantle.
\end{itemize}

\begin{figure}[t]
\begin{center}
\includegraphics[width=14.0cm, height=12cm]
{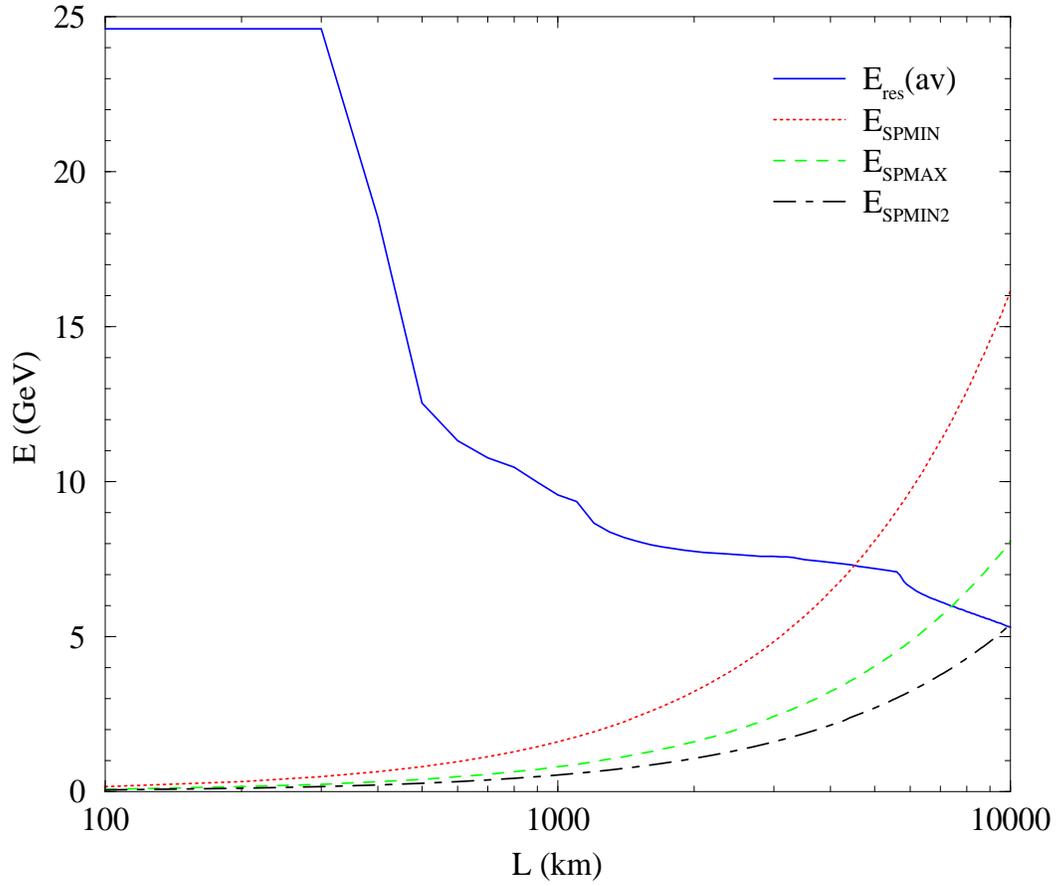}
\caption{\label{fig:eres}
The average $E_{res}$ (blue solid line)
as a function of the baseline length in 
earth. 
Also shown are $E_{\rm SPMIN1}$ (red dotted line), 
$E_{\rm SPMAX}$ (green dashed line) and $E_{\rm SPMIN2}$ 
(black dot-dashed line) in vacuum. They curve owing to the 
logarithmic scale of the horizontal axis.
}
\end{center}
\end{figure}

For a given neutrino energy $E$, 
matter effects begin to be significant only when the neutrinos 
pass through matter densities which are close to their resonant value.
Lower energy neutrinos need higher matter densities to 
hit the resonance condition and vice versa, 
as can be seen from Eq. (\ref{eq:eres}). Since, for upward going
atmospheric neutrinos, the average 
density of the earth increases as $L$ increases, $E_{res}(av)$ 
of Eq. (\ref{eq:eres}) decreases with $L$. 
In Fig. \ref{fig:eres} we show the resonance energy $E_{res}(av)$,
calculated from Eq. (\ref{eq:eres})
with the benchmark (Table \ref{tab:benchmark}) 
values of $\sin^22\theta_{13}$ and
$\Delta m^2_{31}$,
as a function of the distance $L$ travelled by an atmospheric $\nu_\mu$
inside the earth. For each $L$, the average earth matter density has
been calculated using the PREM density profile.
For $L < 1000$ km, the resonance energy $E_{res}$ clearly exceeds 9 GeV. 
Since the flux 
of multi-GeV atmospheric neutrinos falls very fast 
\cite{honda3d} with energy and 
since we restrict our analysis to  
fully contained neutrino events in the detector with $E<11$ GeV, 
we do not expect much matter effect for these short baselines.

\begin{figure}[t]
\begin{center}
\includegraphics[width=16.0cm, height=12cm]
{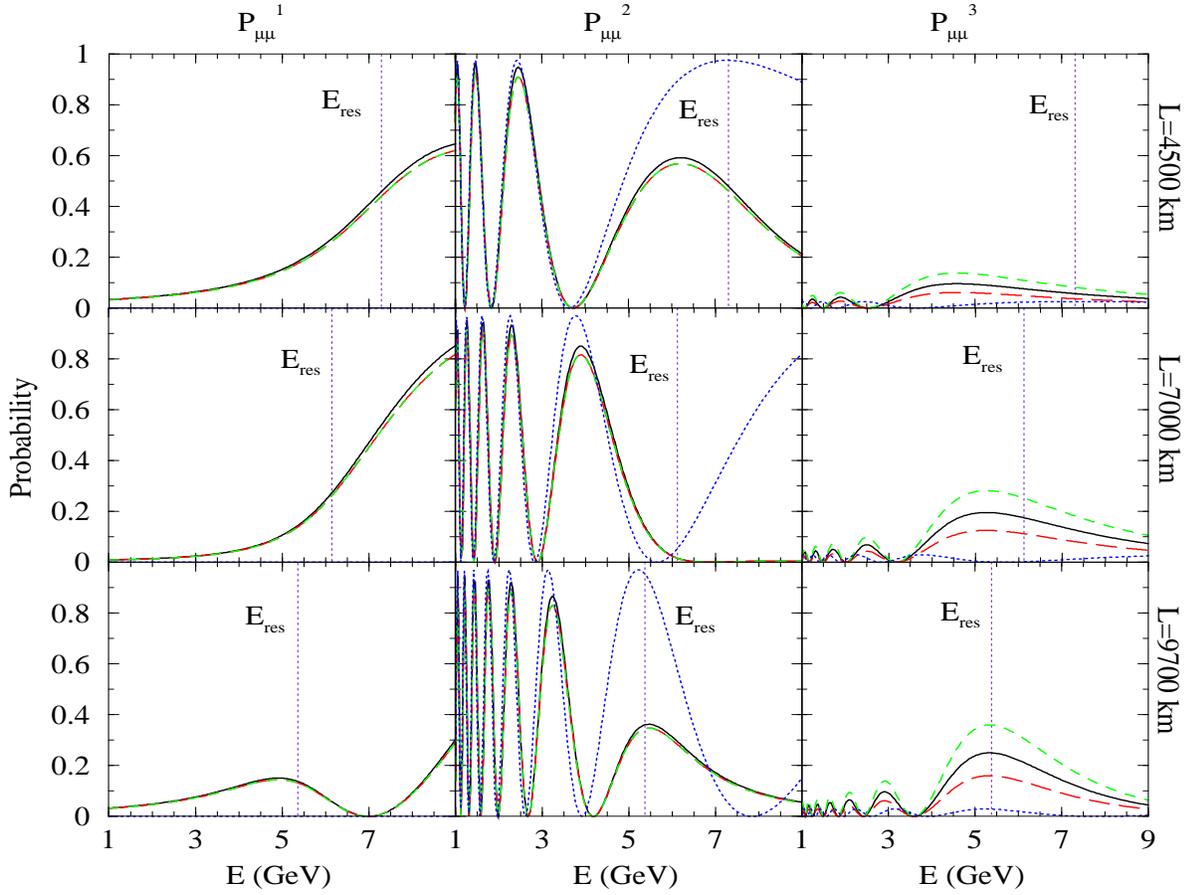}
\caption{
Plots of $P_{\mu\mu}^{1,2,3}$, the notional quantities of 
Eq. (\ref{eq:p22constrho})
defined in the limit $\alpha \rightarrow 0$, 
as functions of the neutrino energy $E$ with a fixed baseline length  
$L$ for different $L$-values as shown.  
The black solid lines, red long-dashed lines and green dashed lines 
correspond to $\nu_\mu$'s travelling in matter
with $\sa=0.5$, 0.4 and 0.6 respectively and the blue dotted lines 
to the same in vacuum with $\sa=0.5$.
}
\label{fig:p223terms}
\end{center}
\end{figure}

\begin{figure}[t]
\begin{center}
\includegraphics[width=16.0cm, height=12cm]{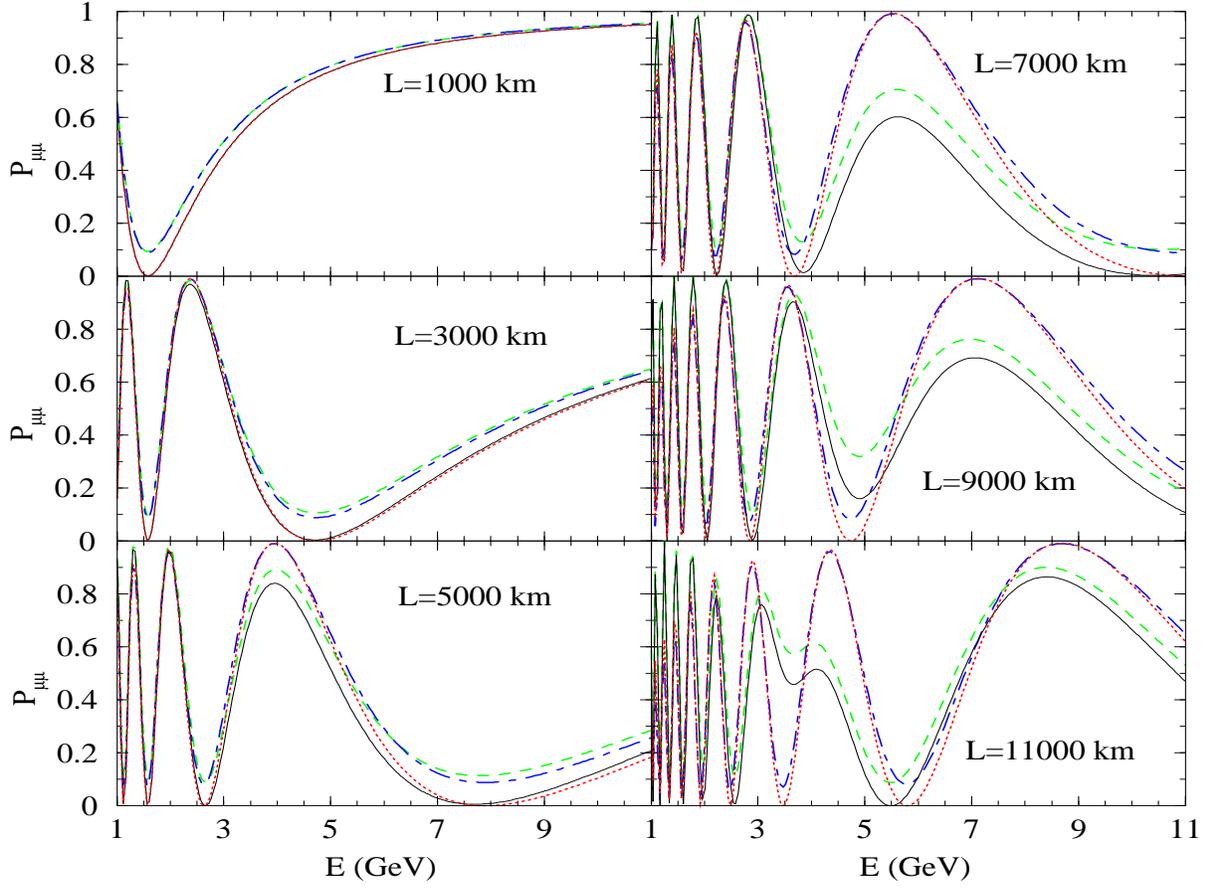}
\caption{
$P_{\mu\mu}$ as a function of 
$E$ for fixed values of $L$.
The black solid and the dashed green lines 
are for neutrinos travelling in matter with $\sa=0.5$ and 
$\sa=0.36$ respectively. The red dotted and the 
blue dot-dashed lines show the corresponding respective cases in vacuum. 
}
\label{fig:probfixedL}
\end{center}
\end{figure}

In order to understand the extent of matter effects for $L > 1000$ km,
we need to take into account another factor apart from the difference
between $E$ and $E_{res}$.  That is the role of the $\ma$-dependent
oscillatory terms in Eq. (\ref{eq:p22constrho}).  
In this context, it is instructive to
start by looking at the extrema of the oscillatory term $\sin^2 (1.27
\ma L/E)$, where $\ma$ is in $eV^2$, $L$ is in km and $E$ in
GeV. We define $E_{\rm SPMIN1}$, $E_{\rm SPMAX}$ and 
$E_{\rm SPMIN2}$ as the respective
values of the energy corresponding to the first\footnote{This means
first or second (as the case may be) starting from the {\it left} in
Fig. \ref{fig:p22fixedE}, but first or second starting from the 
{\it right} in Figs. \ref{fig:p223terms} and \ref{fig:probfixedL}.}
minimum, the first maximum and the second minimum in the
survival probability $P_{\mu\mu}$ as $L/E$ increases.  In other words,
these respectively correspond to the first maximum, the first minimum
and the second maximum as $L/E$ increases in the flavor conversion
probability ${\displaystyle\sum_{\beta = e,\tau}} P_{\mu \beta}$.
Their values in vacuum are, 
\be
E_{\rm SPMIN1} = {2\over\pi} (1.27) \ma L,
\ee
\be
E_{\rm SPMAX} = {1\over\pi} (1.27) \ma L,
\ee
\be
E_{\rm SPMIN2} = {2\over3\pi} (1.27) \ma L,
\ee
are plotted, along with $E_{res}$, against $L$ in Fig. \ref{fig:eres}.  
We deduce
from this figure that $E_{res} \simeq E_{\rm SPMIN1}$ for $L \simeq 4500$
km, while $E_{res} \simeq E_{\rm SPMAX}$ for $L \simeq 7400$ km 
and $E_{res} \simeq E_{\rm SPMIN2}$ for $L
\simeq 9700$ km.  Concentrating on the region 1000 km $< L <$ 4000 km,
we find that here 7.4 GeV $< E_{res} <$ 9.6 GeV, and $E_{res} \gg
E_{SPMIN1}$ so that once again the survival probability $P_{\mu\mu}$
is not much affected by earth matter.  In contrast, for $L > 4000$ km
and with $E$ between 5 and 7 GeV, large matter effects ensue, as
evident from Fig. \ref{fig:p22fixedE}.

The three terms $P^1_{\mu\mu}, \ P^2_{\mu\mu}$ and $P^3_{\mu\mu}$, as
defined in Eq. (\ref{eq:p22constrho}) and given analytically in 
Eqs. (\ref{eq:p22term1})--(\ref{eq:p22term3}), are shown
in Fig. \ref{fig:p223terms} 
as functions of the neutrino energy $E$ for the three special
baseline lengths mentioned above: $L = 4500$ km (upper panels), 7000
km (middle panels) and 9700 km (lower panels).  The dotted blue lines
in Fig. \ref{fig:p223terms} 
show these three terms in vacuum for $\sin^2 \theta_{23} =
1/2$, while the other lines show them in the presence of matter for
different values of $\sin^2\theta_{23}$, as explained in the caption.
For these plots we have taken $\Delta m^2_{21} = 0$, while the other
parameters are as in Fig. \ref{fig:p22fixedE}.  We
discuss below the behavior of the three terms $P^{1,2,3}_{\mu\mu}$
separately\footnote{For a similar discussion on the dependence 
of matter effects in $\pmm$ on $L$ and $E$ see \cite{pomita}.}. 

\begin{enumerate}
\item
$P^1_{\mu\mu}$ is proportional to $\sin^2 \theta^M_{13}$,
cf. Eq. (\ref{eq:p22term1}).  
Though, approximately equal to $\sin^2\theta_{13}$ for
$E \ll E_{res}$, $\sin^2 \theta^M_{13}$ increases with $E$ and reaches
the value $1/2$ at $E = E_{res}$.  Beyond $E_{res}$, $\sin^2
\theta^M_{13}$ continues to rise with $E$ till it saturates to the
maximum value of unity for $E \gg E_{res}$.  For $L = 4500$ km, one
has $S_- \equiv \sin^2 [\{\ma (A+1) - (\ma)^M\}L/8E] \sim 1$ 
for most part of the energy range in Fig. \ref{fig:p223terms}.  
However, for $L = 7000$ km, $S_-$
is $\sim 0.2$ in magnitude at $E = 1$ GeV and rises monotonically to
$\sim 0.9$ at $E = 9$ GeV.  As a result, for both baseline lengths of
4500 km and 7000 km, $P^1_{\mu\mu}$ increases monotonically with
energy.  For $L = 9700$ km, $S_-$ is $\sim 0.8$ around $E = 1$ GeV,
but falls to almost zero at $E = 7$ GeV and rises yet again to $\sim
0.3$ at $E = 9$ GeV.  The consequence is that $P^1_{\mu\mu}$, which is
a product of a continuously rising $\sin^2\theta^M_{13}$ and the
oscillatory $S_-$ factor, has a complicated behavior as a function of
energy at $L = 9700$ km.  We recall here that $P^1_{\mu\mu} = \Delta
P^1_{\mu\mu} \equiv P^1_{\mu\mu} - P^{1,vac}_{\mu\mu}$ is a
pure matter effect term and hence always positive.

\item
The second term $P^2_{\mu\mu}$ is the dominant term in vacuum.  For $E
\ll E_{res}$, there is very little matter effect in it and
$P^2_{\mu\mu}$ remains close to its vacuum value.  As $E$ increases
towards $E_{res}$ and beyond, $\cos^2 \theta^M_{13}$, which appears as
a factor in $P^2_{\mu\mu}$, decreases towards 0.5 at $E_{res}$ and
eventually to near-zero for $E \gg E_{res}$.  For $E < E_{res}$ and
all values of $L$, the matter-dependent oscillatory 
factor $S_+ \equiv \sin^2 [\{\ma (A+1) + (\ma)^M\}L/8E]$ 
closely follows the behavior of the corresponding $\sin^2(\ma L/4E)$
term in vacuum.  But the situation changes drastically for $E >
E_{res}$.  Now the said $S_+$ factor is almost unity at $L = 4500$ km
and 9700 km, but nearly vanishes at $L = 7000$ km.  Thus we can
describe all the cases for $P^2_{\mu\mu}$ in the following way.  While
$P^2_{\mu\mu}$ follows its vacuum oscillation pattern for $E <
E_{res}$, it falls to almost zero as $E$ increases beyond $E_{res}$.
The latter behavior ensues at $L = 4500$ km and 9700 km due to the
fall of $\cos^2 \theta^M_{13}$ to zero, whereas at $L = 7000$ km it is
caused by $S_+$ nearly vanishing.  For $E \sim E_{res}$, $\Delta
P^2_{\mu\mu} \equiv P^2_{\mu\mu} - P^{2,vac}_{\mu\mu}$ is positive at
$L = 7000$ km but negative at $L = 4500$ and 9700 km.

\item
Turning to $P^3_{\mu\mu}$, we see from Eq. (\ref{eq:p22term3}) that its
proportionality to $\sin^2 2\theta^M_{13} \sin^2 [(\ma)^M L/4E]$ makes
it very small in vacuum owing to the smallness of $\sin^2
2\theta_{13}$.  In matter, however, the growth of $\sin^2
2\theta^M_{13}$ with energy (for $E \lsim E_{res}$) till its maximum
value at $E = E_{res}$ makes $P^3_{\mu\mu}$ increase appreciably in
this energy range.  For our example cases shown in Fig. 
\ref{fig:p223terms}, ($E \simeq 5$ GeV
and maximal $\theta_{23}$), it increases from about 0.01, 0.0 and
0.025 in vacuum at $L = 4500,7000$ and 9700 km respectively to about
0.14, 0.20 and 0.25 in earth matter.  This means that, at $E \sim
E_{res}$, $\Delta P^3_{\mu\mu} \equiv P^3_{\mu\mu} -
P^{3,vac}_{\mu\mu}$ is in general positive for all $L$.
\end{enumerate}

The impact of earth matter effects is largest for any $L$ at $E =
E_{res}$ when the sum of the three terms $P^1_{\mu\mu}, P^2_{\mu\mu}$
and $P^3_{\mu\mu}$ becomes most different from its value in vacuum.
In fact, $P^{1,vac}_{\mu\mu}$ $(P^{3,vac}_{\mu\mu})$ is exactly (nearly)
zero for all $E$, while -- for $E \sim 5-7$ GeV --
$P^{2,vac}_{\mu\mu}$ yields SPMIN1 at $L = 4500$ km, SPMAX at $L =
7000$ km and SPMIN2 at $L = 9700$ km.  While matter effects raise
$P^{1,3}_{\mu\mu}$ for all $L$ and $E$ (cf. Fig. \ref{fig:p223terms}), 
apart from mildly
increasing $P^2_{\mu\mu}$ at SPMAX for $L = 7000$ km, they decrease
it when $L = 4500 \ ({\rm SPMIN2})$ and 9700 km (SPMIN2).  There is thus
an important difference between the SPMAX and the two SPMIN cases:
matter effects come with the same relative sign for all the three
terms at SPMAX, but increase $P^{1,3}_{\mu\mu}$ and decrease
$P^2_{\mu\mu}$ at SPMIN1 and SPMIN2.  The SPMAX case at $L =
7000$ km thus exhibits the largest effects of earth matter in the
total survival probability $P_{\mu\mu}$ for $E \sim 5-7$ GeV.  In
contrast, these effects partially cancel in the two SPMIN cases
between the increase in $P^{1,3}_{\mu\mu}$ and the decrease in
$P^2_{\mu\mu}$.  For $L = 4500$ km (SPMIN1), this cancellation is
particularly important and the total $P_{\mu\mu}$ has very little
residual matter effect. Therefore we shall ignore SPMIN1 in our
subsequent matter effect discussion.  For $L = 9700$ km (SPMIN2),
the cancellation is not complete and appreciable matter effects do 
persist in $P_{\mu\mu}$.  As noted earlier, one characteristic of the
SPMAX case ($L = 7000$ km and $E \sim 5$ GeV) is the decrease in
$P_{\mu\mu}$ from its maximal value of near-unity in the presence of
matter since all the three terms $P^{1,2,3}_{\mu\mu}$ increase.  In
contrast, $P_{\mu\mu}$ increases on account of matter in the SPMIN2
case ($L = 9700$ km, $E \sim 5$ GeV, as exemplified here) because of
the dominant decrease in $P^2_{\mu\mu}$.

We can comment on the respective roles of $\theta_{13}$ and 
$\theta_{23}$
in determining the effects of matter on the muon neutrino survival
probability. First, the 
value of $\theta_{13}$, of course, directly controls the extent of
such effects. Figs. \ref{fig:p22fixedE} -- \ref{fig:p223terms} 
have been generated with
$\sin^22\theta_{13} = 0.1$.  For the same $L$ (and hence the same
average matter density), the value of $E_{res}$ is larger for a
smaller $\theta_{13}$, cf. Eq. (\ref{eq:eres}).  
On the other hand, the value of
$(\ma)^M$ is smaller in that case for a constant matter density
independent of $L$.  The result is a larger mismatch between $E_{res}$
(at which $\sin^2 2\theta^M_{13}$ becomes maximal) and the energy
where $P^{2,3}_{\mu\mu}$ reach their extremal values.  Matter effects
in $P_{\mu\mu}$ are therefore reduced for a smaller $\theta_{13}$.
Second,
Fig. \ref{fig:p22fixedE} 
carries evidence of the dependence of the contribution of
matter effects to $P_{\mu\mu}$ on $\sin^2 \theta_{23}$.
The nature and extent of this dependence is more clearly brought out
by Eqs. (\ref{eq:p22term1}--\ref{eq:p22term3}) 
and Fig. \ref{fig:p223terms}.  
While $P^{1,2}_{\mu\mu}$ are seen to be
proportional to $\sin^2 2\theta_{23}$, the third term $P^3_{\mu\mu}$
goes as $\sin^4 \theta_{23}$.  A noteworthy point in this connection
is that, when $\theta_{23}$ is close to being maximal, any change in
its value affects $\sin^2 2\theta_{23}$ much less than $\sin^2
\theta_{23}$.  For instance, when $\theta_{23}$ decreases from
$45^\circ$ to $40^\circ$, $\sin^2 2\theta_{23}$ changes from 1 to
0.97, i.e. by 3\%, while $\sin^2\theta_{23}$ is reduced by 16\% from
0.5 to 0.41.  This is reflected in essentially no visible change in
$P^{1,2}_{\mu\mu}$ when $\sin^2 \theta_{23}$ is
reduced\footnote{$\sin^2 2\theta_{23}$ is the same for $\sin^2
\theta_{23} = 0.4$ or 0.6.} (increased) from 0.5 to 0.4 (0.6),
cf. Fig. \ref{fig:p22fixedE}.  
On the other hand, $P^3_{\mu\mu}$ does increase
appreciably as $\sin^2 \theta_{23}$ is increased from 0.4 to 0.5
leading to an increased $\theta_{23}$-sensitivity in a muon neutrino
disappearance experiment on account of matter effects.  As noted
earlier, the net decrease (increase) of the total $\nu_\mu$ survival
probability $P_{\mu\mu}$ due to matter effects at SPMAX (SPMIN2)
increases (decreases) as the value of $\sin^2 \theta_{23}$ is
increased.  This behavior can now be understood from the dependence of
$P^3_{\mu\mu}$ on $\sin^2 \theta_{23}$.  Let us first consider
SPMAX.  Since the increase due to matter effects in $P^3_{\mu\mu}$
causes a decrease in the total survival probability $P_{\mu\mu}$ at
SPMAX and since $P^3_{\mu\mu}$ in matter increases as
$\sin^2\theta_{23}$ increases, the matter-induced decrease of
$P^{\mu\mu}$ at SPMAX increases with increasing $\sin^2
\theta_{23}$.  For the SPMIN2 case, the dependence of $P^3_{\mu\mu}$
on $\sa$, as discussed above, is more subtle.  In reality, the
main role of the increase in $P^3_{\mu\mu}$ due to matter is to wash
out the corresponding decrease in $P^2_{\mu\mu}$.  Therefore, the
matter-induced relative increase of the survival probability
$P_{\mu\mu}$ at SPMIN2 decreases as $\sin^2 \theta_{23}$ increases.

It is pertinent to stress a couple of points here.  
First, the strong dependence on
$\sin^2 \theta_{23}$ of the total survival probability $P_{\mu\mu}$ in
matter coming from $P^3_{\mu\mu}$, as discussed above, implies that
observed atmospheric $\nu_\mu,\bar\nu_\mu$ events could be used not
only to probe $|D| \equiv |1/2 - \sin^2 \theta_{23}|$ but also the
sign of $D$ to
determine if $\theta_{23} < \pi/4 \ {\rm or} \ > \pi/4$, thus
resolving the ambiguity regarding the octant of $\theta_{23}$.  The
second issue is the effect of earth matter on the shape of the survival
probability as a function of the neutrino energy. 
This is
shown in Fig. \ref{fig:probfixedL} 
for six different choices for $L$.  The various
choices of $\sa$ are given in the figure caption while the
other parameters are the same as in Fig. \ref{fig:p22fixedE}. 
This figure again tells us that matter effects lower
(raise) $P_{\mu\mu}$ from its vacuum values at SPMAX (SPMIN1 and
SPMIN2), resulting in a relative change of the shape of 
$\pmm$ as a function of $E$.  
These matter-dependent effects will of
course be projected onto the observed atmospheric muon neutrino
spectrum which will come folded with $P_{\mu\mu}$.  A detector
with good energy resolution should be sensitive to the spectral shape
of atmospheric muon neutrinos and hence should pick out the shape
distortion owing to matter effects.

\section{\label{sec:updown}
Up-down asymmetry in an ICAL-like detector}

A very powerful variable, which clearly displays  
oscillation effects for 
atmospheric neutrinos (antineutrinos), is 
the up-down ratio \cite{skatmud} $U_N/D_N$ ($U_A/D_A$). 
Here $U_N$ ($U_A$) is the number of events recorded for the
``upward'' muon neutrinos (antineutrinos) 
coming with $\cos\xi < 0$ 
and $D_N$ ($D_A$) is the number of 
atmospheric events recoded for the 
``downward'' muon neutrinos (antineutrinos) with $\cos\xi > 0$, 
$\xi$ denoting
the zenith angle of the $\numu$/$\anumu$ trajectory.
Downward (upward) neutrinos and antineutrinos cover short (long) 
distances. 
Hence downward ones hardly undergo any flavor 
transformation while upward ones are subjected to full 
flavor oscillations, the relevant oscillation lengths being 
in the range $10^2-10^3$ km. 
Thus
the up-down ratio $U/D$ yields crucial information 
on the degree of oscillations for upward neutrinos and antineutrinos.
Being relatively insensitive\footnote{We have more
to say on this issue later in this section.}
to the uncertainties in the 
atmospheric neutrino flux calculations, it is a reliable 
measure of the survival probability $\pmm$. As such, it can be 
used to study both the vacuum oscillations of muon neutrinos
and the effect of matter on them \cite{indu}.
We shall consider the above ratio for atmospheric muon neutrino 
and antineutrino events in a large magnetized iron calorimetric
detector.
Such a detector named ICAL, has been conceived in the 
proposed underground neutrino laboratory INO \cite{ino}
in India.
While the initial proposal is for a 
detector mass of 50 kton for ICAL,
the aim is to enlarge it to a final detector size of 100 kton.
The proposed detector would have a modular structure with stacks 
of $\sim 6$ cm thick magnetized iron plates interleaved with 
$\sim 2.5$ cm of resistive plate chambers (RPC) made of glass 
as the active detector material. A uniform magnetized field of 
$1-1.4$ Tesla would enable the charge discrimination of muons
distinguishing between $\numu$-- and $\anumu-$induced events.
The full detector would be divided into three modules, the modular 
structure allowing the start of detector operations with those 
modules that are ready even when others are still under construction.
Further details are given in \cite{ino}.

The expected number of $\mu$ (or $\bar\mu$) 
events, induced by oscillating 
atmospheric muon neutrinos (or antineutrinos) and 
recorded in a detector such as ICAL, is 
given by
\begin{eqnarray}
N_l
= 
&n_T &
\int_{\xi_{min}}^{\xi_{max}} d\cos \xi
\int_0^\pi d\xi^\prime \tilde R(\xi,\xi^\prime)
\int^{E_{max}}_{E_{min}} dE
\int_0^\infty dE^\prime \tilde R(E,E^\prime)
~~ \sigma_l (E^\prime)~ \epsilon_l(E^\prime)
\nonumber\\
&\times&\left[
{d^2\phi_l (E^\prime ~,\xi^\prime ~) \over dE^\prime ~~
d\cos\xi^\prime ~}
{\ }P_{l l} (E^\prime, \xi^\prime)
+
{d^2\phi_{l^\prime} (E^\prime ~,\xi^\prime ~) 
\over dE^\prime ~d\cos\xi^\prime}
{\ }P_{{l^\prime} l} (E^\prime, \xi^\prime)
\right]
,
\label{Eq:rate}
\end{eqnarray}
where, $E^\prime$ and $E$ are the true and 
experimentally reconstructed neutrino energies 
respectively and $\tilde R(E,E^\prime)$ 
is the energy resolution function of the detector.
Likewise, $\xi^\prime$ and $\xi$ are the true and 
reconstructed neutrino zenith angle and $R(\xi,\xi^\prime)$ 
is the zenith angle resolution function of the detector. 
It is convenient for us to adopt\footnote{
We have checked that for the $E$ and $L$ 
bin sizes used in this paper
the resolution functions make only little difference to the final 
results. Hence, the precise form of the resolution 
function  is unimportant.}
Gaussian forms for the
energy and length (zenith angle) resolution functions of the detector:
\be
\tilde R(x,x^\prime) = 
\frac{1}{\sqrt{2\pi}\sigma_{x^\prime}}\exp\left(\frac{-(x-x^\prime)^2}
{2\sigma_{x^\prime}^2}\right)~,
\label{eq:resol}
\ee
where $x$ is $E$ or $\xi$ ($L$).
For the energy resolution we assume $2\sigma_{E^\prime} =0.3 E^\prime$. The
detector resolution for the zenith angle and hence the distance travelled
by the neutrino is expected to be somewhat better and we assume 
$2\sigma_{L^\prime}= 0.2 L^\prime $. 
These numbers are more or less the same as the full
widths 
of the nonGaussian resolution functions
given in \cite{ino}.
The total distance
$L$ and the zenith angle $\xi$
are related by
\be
L= \sqrt{(R_e+h)^2 - {R_e}^2 \sin^2 \xi} - R_e \cos \xi~,
\ee
where $R_e$ is the radius of the earth and $h$ the height of 
the atmosphere.
Among the rest of the quantities of 
Eq. (\ref{Eq:rate}), $n_T$ denotes the number of target nucleons
times the total live time of the detector,
$\sigma_l(E^\prime)$ is the total $\nu_l N \rightarrow l X$ 
scattering cross-section of a $\nu_l$ of energy $E^\prime$ and 
$\epsilon_l(E^\prime)$ is the trigger efficiency of the 
magnetized calorimeter.
Here the index $l$ can be $\mu$ or $\bar\mu$
and $d^2\phi_l /dE^\prime d\cos\xi^\prime$ is the differential 
flux of atmospheric neutrinos $\nu_l$. 
Finally, $P_{ll}$ gives the survival 
probability for the atmospheric $\numu$ (or $\anumu$) 
and $P_{{l^\prime} l}$ the transition probability of $\nu_l^\prime$
to $\nu_l$, $l^\prime$ being either $e$ or $\bar e$
correlated to $l$ being $\mu$ or $\bar\mu$.

We shall present our results for a
100 kT of active detector mass and ten years of running,
assuming a conservative 50\% trigger efficiency\footnote{
 In reality, this could be as high as 80\%, 
S. R. Dugad, private communication.}
for the detector
\footnote{For any other 
detector mass and efficiency, the statistics can be  
accordingly scaled.}. 
In order to calculate the number of detected events,
we use the latest three-dimensional
atmospheric neutrino fluxes provided 
by Honda {\it et. al.} \cite{honda3d}, which is also used by the 
INO collaboration \cite{ino}.
For the reaction cross-section 
we consider only the DIS process and 
use the cross-sections given by the CTEQ collaboration \cite{ctEq}.
We then distribute the number of $\mu^-$ ($\mu^+$) events induced by the 
$\numu$ ($\anumu$), into various zenith angle and 
energy bins. In the range $E=1-11$ GeV, the events are
divided into 
five energy bins, each of width 2 GeV. The zenith angle binning is 
presented in Table \ref{tab:zen}. The bins $1-6$ ($7-12$)
correspond to $\cos\xi >0$ ($<0$) 
and hence contain downward (upward) going neutrinos/antineutrinos.
Upward going ones have been binned according to $L_m$,
the distance they travel within earth matter. 
The range of $L_m$ (in km),
corresponding to each zenith angle bin, is 
given in Table \ref{tab:zen}.
The bins for downward neutrinos are arranged to have a 
structure similar to that of upward neutrinos.
We simulate the ``data'' for the prototype detector,
assuming certain ``true'' values 
for the oscillation parameters namely those benchmarked in 
Table \ref{tab:benchmark}, and 
distribute them into the five energy bins and the twelve zenith angle bins.
Thus we have $5\times 12=60$ bins of data for the muon events. 
The iron detector we consider is magnetized 
and therefore has charge discrimination capability.  So  
we have 60 bins of $\mu^-$ data for the neutrino
channel and another 60 bins of $\mu^+$ data for the antineutrino channel.
Our total data set therefore comprises 120 bins.

\begin{figure}[t]
\begin{center}
\includegraphics[width=16.0cm, height=12cm]{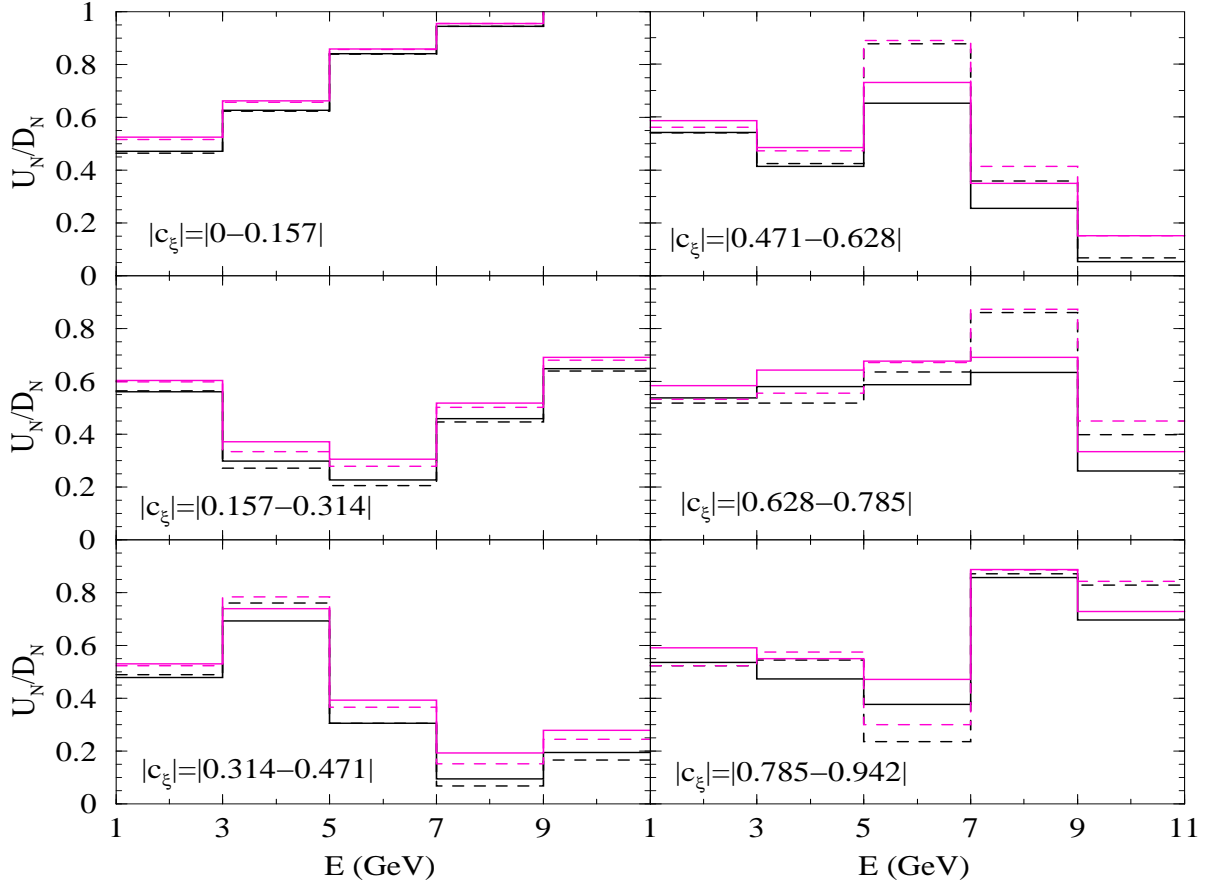}
\caption{
The up-down asymmetry expected for muon neutrinos in energy 
bins of width 2 GeV. 
The six panels show the data for 
six different zenith angle
(or $L$) 
bins corresponding to upward neutrinos travelling 
between $L_m=0-2000$ km, $L_m=2000-4000$ km, $L_m=4000-6000$ km,
$L_m=6000-8000$ km, $L_m=8000-10000$ km and 
$L_m=10000-12000$ km respectively inside the earth, cf. Table 
\ref{tab:zen}.
The solid black lines and the solid magenta lines are for 
neutrinos travelling in matter with $\sa=0.5$ and 0.36 respectively.
The dashed black lines and the dashed magenta lines are for 
neutrinos travelling in vacuum with $\sa=0.5$ and 0.36 respectively.
For all cases we have taken $\ma=2\times 10^{-3}$ eV$^2$ and 
the benchmark 
values of Table \ref{tab:benchmark} for the other parameters.
}
\label{fig:updown}
\end{center}
\end{figure}

The up-down ratio $U_N/D_N$ for $\numu$'s is shown in 
five different energy bins in Fig. \ref{fig:updown}. 
However, 
instead of integrating over all upward and downward zenith 
angles, we have chosen to divide the said 
ratio into six zenith angle bins.
Each of the panels in Fig. \ref{fig:updown} contains $U_N/D_N$
for a certain range of the modulus of the 
zenith angle as shown in the legend.
For instance, the first (last) panel on the top left
(bottom right) contains the ratio $U_N/D_N$ for 
$0 \leq |\cos\xi| \leq 0.157$ ($0.785 \leq |\cos\xi| \leq 1$)
and calculated by taking upward going neutrinos with 
$-0.157 \leq \cos\xi \leq 0$ ($-1\leq \cos\xi \leq -0.785$)
and downward going neutrinos with 
$0 \leq \cos\xi \leq 0.157$ ($0.785\leq \cos\xi \leq 1 $).
The black and magenta solid lines in Fig. \ref{fig:updown} 
describe the $U_N/D_N$ spectrum for the realistic case of upward
neutrinos travelling through earth matter and with 
$\sa=0.5$ and $\sa=0.36$ respectively. In contrast, the 
similar dashed lines show the hypothetical
$U_N/D_N$ spectrum for the corresponding values of $\sa$,
if the upward neutrinos were assumed to be 
travelling in vacuum even inside the earth. 
For the same value of $\sa$,
a comparison of the solid line with
the corresponding dashed line 
shows the effect of earth matter in 
changing the muon neutrino up-down ratio.
Furthermore, the 
degree of this change is seen to 
depend on the value of $\sa$. 

\begin{figure}[t]
\begin{center}
\includegraphics[width=16.0cm, height=12cm]
{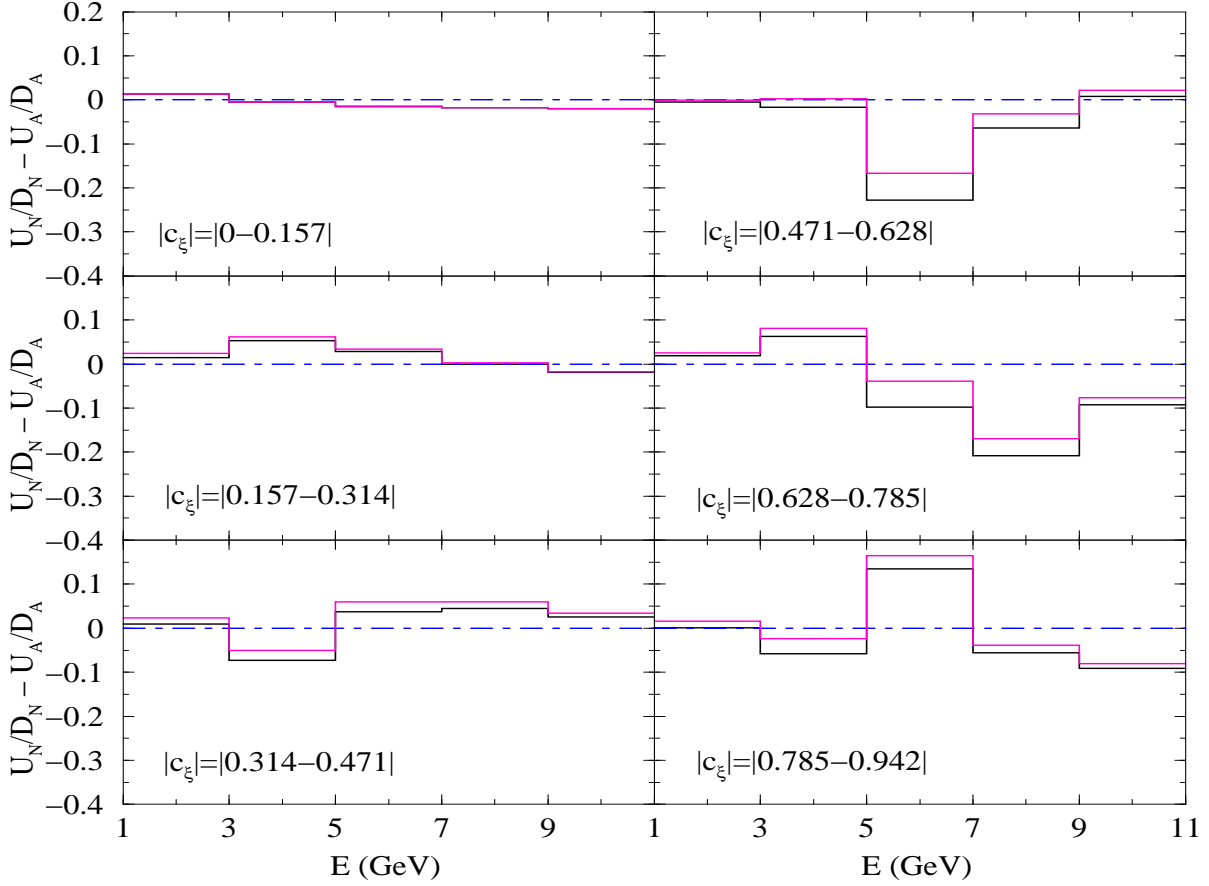}
\caption{
The difference between the 
up-down ratio for the neutrinos ($U_N/D_N$) and 
antineutrinos ($U_A/D_A$)
shown for the various energy and zenith angle bins. 
The solid black lines are for neutrinos/antineutrinos 
travelling in matter 
with $\sa=0.5$, while the solid magenta lines are 
neutrinos/antineutrinos travelling in matter 
with $\sa=0.36$. The dot-dashed blue line 
shows $U_N/D_N - U_A/D_A =0$ for reference.
The other 
oscillation parameters are chosen as in Fig. \ref{fig:updown}.
}
\label{fig:updowndiff}
\end{center}
\end{figure}

An even better way of bringing out  
matter effects\footnote{For a normal (inverted) mass ordering, matter
effects at baseline lengths $L > 1000$ km
for $\nu_\mu$'s ($\bar\nu_\mu$'s) are significantly larger
than for $\bar\nu_\mu$'s ($\nu_\mu$'s), cf. discussion at the end of
the section.} in atmospheric muon neutrinos
is to look at the difference $U_N/D_N - U_A/D_A$, where the subscript
$A$ now refers to $\bar\nu_\mu$'s.  
The up-down ratio
$U/D$ yields $\pmm$ which is the same for neutrinos and  
antineutrinos in vacuum so long as CPT is conserved.  Therefore, the 
difference in the up-down ratio between neutrinos and antineutrinos
gives a direct measure of the matter effects in $\pmm$.
In fact, it relates to $\Delta P_{\mu\mu}$ of Eq. (\ref{Eq:deltapmm}).
Fig. \ref{fig:updowndiff} shows our 
results for the same choice of zenith angle and energy bins and 
for the same set of oscillation parameters with an assumed normal mass 
ordering as made in Fig. \ref{fig:updown}.
Substantial matter effects in terms of the deviation of the difference
$U_N/D_N - U_A/D_A$  
from zero are concretely shown in Fig. \ref{fig:updowndiff}
and in a 
more pronounced manner in the right hand panels.
Once again, the strength of the matter part in the 
survival probability is seen to depend on the value of $\sa$.
\\

\noindent
We can summarize the main characteric 
features of 
Figs. \ref{fig:updown} and \ref{fig:updowndiff} 
as follows:
\begin{itemize}
\item For a normal mass ordering ($\ma >0$), 
atmospheric muon neutrinos 
undergo significant matter effects for $E\gtap 3$ GeV 
and $L\gtap 4000$ km. A similar statement holds true for 
muon antineutrinos if $\ma < 0$.
\item Matter effects increase (decrease) 
the ratio $U_N/D_N$ in the $E$ and 
$\cos\xi$ bins corresponding to the SPMINs (SPMAX)
for a normal mass ordering. For an inverted mass
ordering, a similar statement can be made about $U_A/D_A$.
\item The largest impact of earth matter comes in the bin 
$E=5-7$ GeV and $-0.628 \leq \cos\xi \leq -0.471$ 
($6000 \leq L_m\leq 8000$ km),
corresponding to the SPMAX. 
Here $U_N/D_N - U_A/D_A$ reduces for a 
normal mass ordering 
to $-0.3$ to $-0.25$, 
depending on the value of $\sa$.
\item The dependence of the strength of matter effects on the value of 
$\sa$ is most clearly brought out in the 
energy and zenith angle bin mentioned in the preceding bullet.  
Reducing $\sa$ from the maximal 0.5 to 0.36 brings a nearly 
10\% change in the difference   $U_N/D_N - U_A/D_A$.
This highlights the possible role of matter effects in achieving 
a better sensitivity to $\sa$ and hence a better resolution of  
its difference from maximality.
\end{itemize}

Let us make two further comments. First, 
the main advantage of the up-down asymmetry parameter, 
constructed from directly recorded events,
is its insensitivity to the error in 
the absolute normalization of the atmospheric 
neutrino flux, on account of cancellations between the upward 
and downward fluxes.
Recall that this uncertainty has been the main source of systematic 
error in extracting the mass squared difference and mixing parameters from 
any set of observed atmospheric neutrino events.
Additionally, the charge discrimination capability of 
an ICAL type of a detector should be able to bring out 
earth matter effects in the atmospheric neutrino signal 
much more effectively by constructing the difference 
$U_N/D_N - U_A/D_A$ from recorded events.
The second comment relates to our assumption of a normal neutrino 
mass ordering, automatically implying larger  
matter effects in the 
$\numu$ rather than the $\anumu$ channel
for $L > 1000$ km. In case Nature has chosen 
an inverted ordering of the neutrino masses, the above features would 
hold qualitatively except that the situation would be reversed between the 
$\numu$ and $\anumu$. In particular, the major change in the plot of 
Fig. \ref{fig:updowndiff} would be that of a sign.

\begin{table*}
\begin{tabular}{cccc}
\hline
\hline
bin & $\cos\xi$ & Total distance $L$ in km& 
Distance in earth $L_m$ in km \cr
\hline
1 & 1.000      to   0.785 & $15.0-19.1$ & 0\cr
2 & 0.785 to  0.628 & $19.1-23.8$ & 0\cr
3 & 0.628 to  0.471 & $23.8-31.7$ & 0\cr
4 & 0.471 to  0.314 & $31.7-47.3$ & 0\cr
5 & 0.314 to  0.157 & $47.3-91.5$ & 0\cr
6 & 0.157 to  0.000 & $91.5-437.4$ & 0\cr
7 &   0.000     to    -0.157  & $437.4-2091.5$ & $0-2000$\cr
8 &  -0.157 to -0.314 & $2091.5-4047.3$ & $2000-4000$\cr
9  & -0.314 to -0.471 & $4047.3-6031.7$& $4000-6000$\cr
10  & -0.471 to -0.628 & $6031.7-8023.8$& $6000-8000$\cr
11  & -0.628 to -0.785 & $8023.8-10019.1$& $8000-10000$\cr
12  & -0.785 to -1.000 & $10019.1-12757.0$ & $10000-12500$\cr
\hline\hline
\end{tabular} 
\caption{
The zenith angles bins used. 
}
\label{tab:zen}
\end{table*}

\section{\label{sec:chisq}
Details of the Statistical Analysis}

Our twin goal is to check the sensitivity of atmospheric $\nu_\mu$'s
and $\bar\nu_\mu$'s to any deviation of $\theta_{23}$ from its maximal
value $\pi/4$ and to see the effects of earth matter on this
sensitivity.  In order to achieve this end, we perform a statistical
analysis of simulated ``data'' generated in an ICAL-like calorimeter,
assuming certain ``true'' values for the oscillation parameters.
For our error analysis we define a $\chi^2$ parameter
following \cite{maltonipull,lisipull,skthesispull} as
\be
\chi^2_{atm} \equiv \min_{ \xi_k }
\left[\sum_{n=1}^{120}\left(\frac{\tilde{N}_{n}^{theory} - N_n^{data}}
{\sigma_n^{stat}}\right)^2
+ \sum_{k=1}^K\xi_k^2\right]~, 
\label{Eq:chipull}
\ee
In Eq. (\ref{Eq:chipull}) 
$N^{data}_n$ are the observed number of events in bin $n$
and $\sigma^{stat}_n$ are the statistical errors. The systematic errors
in the data and the theory are accounted for through the set of pulls
$\{\xi_k\}$.  The latter are defined in such a way that the number of
expected events $N^{theory}_n$, in bin $n$, corresponds to $\xi_k = 0$
and their $1\sigma$ deviations are given by $\xi_k = \pm 1$.  The
theoretical and experimental uncertainties then cause the expected
number of events to shift to ${\tilde N}^{theory}_n$:
\be
\tilde{N}_{n}^{theory} = N_{n}^{theory} \left[
1+ \sum_{k=1}^K \pi_n^k \xi_k \right] + {\cal O}(\xi^2_k)~,
\label{Eq:rth}
\ee
In Eq. (\ref{Eq:rth}) 
${\tilde N}^{theory}_n$ has been expanded in powers of
$\xi_k$, keeping only linear terms.  The quantities $\pi^k_n$ give the
fractional rate of change of $N^{theory}_n$ due to the $k$th
systematic uncertainty.

The most important theoretical systematic uncertainties come from our
lack of knowledge of the predicted atmospheric neutrino fluxes.  In
order to cover these, we take an absolute normalization error of 20\%,
a ``tilt'' factor of 5\% to account for the spectral uncertainty, 
an uncertainty of 5\% in the neutrino-antineutrino flux ratio 
and an uncertainty of 5\% in the zenith angle dependence. 
For the DIS
cross-sections, we consider an uncertainty of 10\% -- assuming a
modest improvement in our present understanding of these
cross-sections by the time ICAL starts operating.  For the other
experimental systematic uncertainties, we take a 5\% consolidated
error.  We take these somewhat arbitrarily assumed plausible
systematic errors for the proposed experiment, pending actual
estimates to be provided by the INO collaboration in future.
We calculate the uncertainties $\pi_n^k$ in Eq. (\ref{Eq:rth}) 
using an approach similar to that used in \cite{maltonipull}.
Therefore, $\pi_n^k$ is taken as 20\%, 10\% and 5\% of the 
number of events for all bins when 
$k$ corresponds to the absolute flux normalisation, cross-section 
uncertainty and experimental systematic error respectively. 
The $\pi_n^k$s
for $k$ corresponding to the ``tilt'' factor and zenith angle dependence 
of the $\numu$ and $\anumu$ flux
are calculated using the method detailed in \cite{maltonipull}.
For the flux uncertainty coming from neutrino-antineutrino flux ratio, 
we take $\pi_n^k = 2.5\%$ of the number of events 
for all the neutrino bins and  $\pi_n^k = -2.5\%$ of number of events
for all the antineutrino bins.

Unless otherwise stated, for the numerical analysis done in
this paper, the number of ``data'' events $N^{data}_n$ in the $n$th
bin is simulated at ``true'' values\footnote{In what follows, we shall
always distinguish between the ``true'' values of the oscillation
parameters chosen by Nature at which we generate our projected data
and the fitted values of those parameters which are constrained by the
same data.  The true values will henceforth be written as
$\sin^2\theta_{23}$(true) etc., while the fitted values will be
referred to as $\sin^2\theta_{23}$ etc.} of the parameter set $\{p \
({\rm true})\}$, $p$ covering $\Delta m^2_{31}$, $\Delta m^2_{21}$,
$\sin^2\theta_{12}$ and $\delta$, which are set at the corresponding
benchmark values listed in Table \ref{tab:benchmark}.  
The choice
of a normal mass ordering is
for the sake of being definite rather than on account of any
theoretical prejudice. We shall also present some results for the 
inverted mass ordering. 
The true values of the remaining oscillation
parameters $\sin^2\theta_{23}$ and $\sin^2\theta_{13}$ will be varied
to account for their impact on matter effects and will be stated
explicitly as and when used.  (Any change, whenever made in the
assumed true values of the other parameters, will also be mentioned
explicitly.)  For each such simulated data set, we do a statistical
analysis using Eqs. (\ref{Eq:chipull}) and 
(\ref{Eq:rth}) to find the sensitivity of ICAL to
any deviation of $\theta_{23}$ from maximality as well as to the
fixation of its right octant.  ${\tilde N}^{theory}_n$ is calculated
for a given set of oscillation parameters using Eq. (\ref{Eq:rth}). 
The RHS of Eq. (\ref{Eq:chipull}) is then minimized with respect to the
pulls $\xi_k$ to obtain $\chi^2_{atm}$ as a function of the
oscillation parameters.  The latter function is then further minimized
by varying the oscillation parameters within their allowed ranges to
obtain the sensitivity plots for $\theta_{23}$.

The extent of earth matter effects is determined by the true value of
$\sin^2\theta_{13}$.  Therefore, we also include the constraints on
$\sin^2\theta_{13}$ expected to ensue from the combined data expected 
from the next generation reactor
\cite{react13} 
and long baseline accelerator \cite{lbl} 
experiments. So we define the 
{\it combined} $\chi^2$ for the ICAL experiment as
\be
\chi^2_{comb} = \chi^2_{atm} + \left(\frac{\sch - \sch{\rm (true)}}
{\sigma_{s_{13}^2}}\right)^2~.
\label{Eq:chicombined}
\ee 
In Eq. (\ref{Eq:chicombined}), 
$\chi^2_{atm}$ is as given by Eq. (\ref{Eq:chipull}).  Furthermore, the
second RHS term tries to take into account the above-mentioned bounds
on $\sin^2\theta_{13}$ from future laboratory experiments with the
denominator $\sigma_{s^2_{13}}$ denoting the $1\sigma$ uncertainty in
it which is assumed \cite{huber10} to be at the level of 3.5\%.

\section{\label{sec:maxim}
Deviation of $\theta_{23}$ from its maximal value}

\begin{figure}[t]
\begin{center}
\includegraphics[width=16.0cm, height=12cm]
{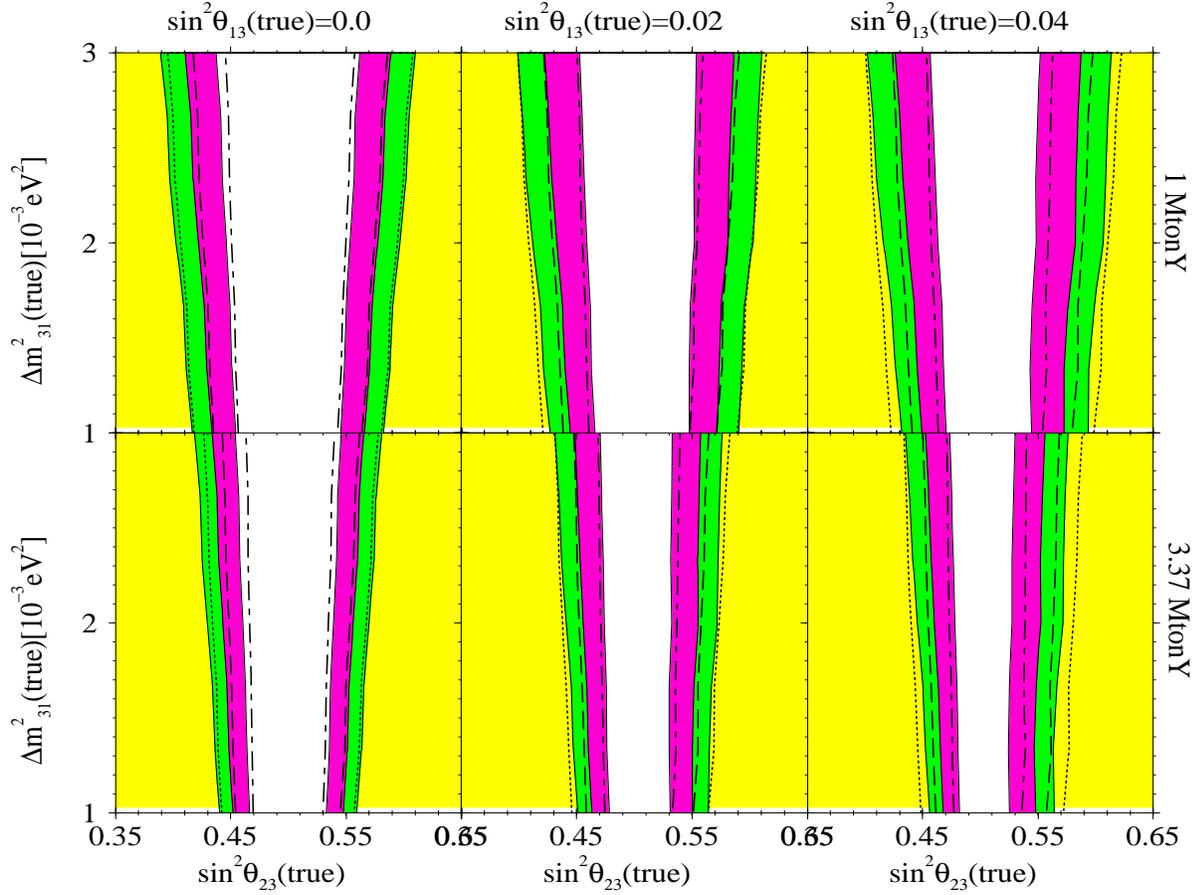}
\caption{
The regions of $\mat$ and $\sat$ where maximal $\theta_{23}$
mixing can be rejected by using 1 MtonY (upper panels) 
and 3.37 MtonY (lower panels) atmospheric neutrino 
data in ICAL
at $1\sigma$ (white band), $2\sigma$ (blue band)
and $3\sigma$ (green band). The hollow dark lines show the 
corresponding bands for neutrinos travelling in pure vacuum.
Benchmark parametric values of Table \ref{tab:benchmark}
have been assumed.
}
\label{fig:sens}
\end{center}
\end{figure}

Let us turn now to the focal point of this paper: how well will the
deviation of $\theta_{23}$, if any, from its maximal value be probed
with atmospheric neutrinos in the foreseeable future?  We take the
combined information from 1 MtonY of simulated data in ICAL, adding
on the constraint on $\sin^2\theta_{13}$ to come from future reactor
and accelerator experiments.  Our procedure is to generate the data at
a certain nonmaximal value of $\sin^2\theta_{23}$(true) and then fit
this data with the maximal muon neutrino mixing angle $\theta_{23} =
\pi/4$, choosing different values of $\sin^2\theta_{23}$(true).  
The results are displayed in Fig. \ref{fig:sens}.
At each point in this $\Delta m^2_{31}$(true) -- 
$\sin^2\theta_{23}$(true) plane, 
we simulate the 120 bin data in 
ICAL, taking $\sin^2\theta_{13}$(true)
$= 0.00$ (left panel), 
$0.02$ (middle panel) and $0.04$ (right panel)
respectively and assuming for all the 
other parameters the benchmark values given in Table 
\ref{tab:benchmark} to be true.
Each such data set is then fitted back  
using Eqs. (\ref{Eq:rth}), (\ref{Eq:chipull}) and (\ref{Eq:chicombined})
with $N_n^{theory}$ calculated for maximal $\theta_{23}$ mixing.
The parameters $\ma$ and $\sch$ are allowed to take any possible 
values in the fitted $N_n^{theory}$,
while $\ms$ and $\sss$ are allowed to vary freely within 
7\% and 15\%
of their assumed true values respectively\footnote{
These are the bounds on the solar parameters expected 
from the future solar and long baseline reactor 
neutrino experiments. See for example
\cite{th12future} for a recent detailed discussion.}.

The upper and lower panels of Fig. \ref{fig:sens}
correspond to respective exposures of 1 MtonY and 
3.37 MtonY\footnote{3.37 MtonY would
roughly be the statistics needed in ICAL to match the number of
fully contained muon events in the experiment SK20 which is
described later in this section.}.
The regions of $\sin^2\theta_{23}$(true) and $\Delta m^2_{31}$(true)
within the white, blue and green bands of Fig. \ref{fig:sens} 
show the true values
of those quantities for which the distinction of a maximal from a true
nonmaximal value of $\theta_{23}$ will not be possible at the
$1\sigma$, $2\sigma$ and $3\sigma$ levels respectively for the
specified exposure.  The broken lines give the corresponding limits of
$\sin^2\theta_{23}$(true) in case earth matter effects were
deliberately switched off by hand.  Among those the long-dashed,
dot-dashed and dotted lines respectively yield the $1\sigma$,
$2\sigma$ and $3\sigma$ limiting values of $\sin^2\theta_{23}$(true).  
A comparison of the broken lines with the corresponding
continuous lines show the following feature.  Matter effects tend to
increase somewhat the sensitivity of ICAL to test the maximality of
$\sin^2\theta_{23}$. 

We present in Table \ref{tab:max} the intervals of $D$(true) $= 1/2 -
\sin^2\theta_{23}$(true) beyond which maximal mixing could be ruled out
at the $3\sigma$ level, with and without matter effects.  The latter
corresponds to a fictitious environment of pure vacuum and has been
included to see the quantitative role of matter effects here.
Specifically, for $\Delta m^2_{31}$(true) $= 2.0 \times 10^{-3} \
{\rm eV}^2$ and $\sin^2\theta_{13}$(true) $=0.04 ~(0.00), \
\sin^2\theta_{23}$(true) can be distinguished by ICAL from 
the maximal value of 0.5
at the $3\sigma$ level 
within 17\% (20\%) from 1 MtonY of exposure and 
within 11\% (14\%) if the statistics was increased to 3.37 MtonY.  
The corresponding ranges of $D$(true) are what Table \ref{tab:max}
lists.  This is comparable to the sensitivity of 
the combined data from the 
forthcoming accelerator-based long baseline experiments\footnote
{The authors of \cite{antusch} use a combination of simulated data 
set from five years of 
running of MINOS, ICARUS, OPERA, T2K and NO$\nu$A 
each, expected to come in the next ten years.}
to a
deviation from maximality of $\sin^2\theta_{23}$, which is 
\cite{antusch} $\sim
14$\% at $3\sigma$ 
(for $\Delta m^2_{31}$(true)$ = 2.5\times 10^{-3}$ eV$^2$)
\footnote
{This estimated uncertainty will
be quite different if the true value of $\Delta m^2_{31}$ deviates
substantially from $2.5\times 10^{-3}$ eV$^2$.}.
Our sensitivity to $D$ is
also comparable to that expected with atmospheric neutrinos in 
very large futuristic water
Cerenkov detectors.  For statistics that is 20 (50) times the
current SK statistics, denoted as SK20 (SK50), a very large water
Cerenkov atmospheric neutrino experiment 
is expected to test a deviation from a maximal
$\sin^2\theta_{23}$ upto \cite{maltonimax23} 23\% (19\%) at $3\sigma$.

\begin{table}
\begin{center}
\begin{tabular}{|l|l|l|}
\hline
\multicolumn{3}{|c|}{Range of $D$(true) in units of $10^{-2}$} \\
\hline
 & \hspace*{2cm} matter & \hspace*{1.5cm} vacuum \\
\hline
$\sin^2\theta_{13}$(true) 
& \ \ \ 0.00 \hspace*{1.2cm} 0.02 \hspace*{1.2cm} 0.04 
& \ \ \ 0.00 \hspace*{1cm} 0.02 \hspace*{1cm} 0.04 \\
\hline
1 MtonY 
& [9.8,-9.7] \ \ \ [8.8,-10.3] \ \ \ [8.4,-10.7] 
& [9.4,-9.4] \ \  [9.2,-10.2] \ \ \ [9.0,-11.1] \\
\hline
3.367 MtonY & [7.0,-7.0]  \hspace*{.3cm} [5.9,-7.2] \hspace*{.6cm}[5.5,-7.2] 
& [6.6,-6.7] \ \ [6.3,-7.3] \ \ \ \ \ [6.0,-8.0] \\
\hline
\end{tabular}
\caption{\label{tab:max}
Simulated ranges of deviation $D$ from maximal $\sin^2\theta_{23}$,
allowed at $3\sigma$, after two different exposure times at ICAL.
Here $\mat$ has been 
taken to be $2.0 \times 10^{-3} $eV$^2$.}
\end{center}
\end{table}

Let us make some observations here.  
While atmospheric neutrino data in
a very large water Cerenkov detector could test 
\cite{maltonimax23} maximality in
$\sin^2\theta_{23}$ almost independently of the values of 
$\scht$ and $\mat$, the 
sensitivity at a detector like 
ICAL to $D$ would depend on both these oscillation 
parameters. 
The dependence of the latter on $\scht$ comes from 
the greater importance of matter effects here. 
The impact of $\mat$ on the measurement $D$ comes from the 
dependence of $\sin^2\theta_{23}$ sensitivity 
to the spectral shape of the data at ICAL, cf. \S 3.
For a larger $\Delta m^2_{31}$(true), 
there is a larger averaging of the oscillation signal and the
$\sin^2\theta_{23}$ sensitivity is reduced slightly.  
Such a fact is evident
from Figs. 8 and 9 which display the muon $(\mu^-)$ zenith angle
spectrum for $\Delta m^2_{31} = 2.0 \times 10^{-3} \ {\rm
eV}^2$ and $4.0 \times 10^{-3} \ {\rm eV}^2$ respectively.  The four
panels in both figures show the zenith angle dependence of the muon
events in four different energy bins which are labeled.  While the
solid black and the magenta dotted lines are for $\sin^2\theta_{23} =
0.5$ (maximal) and 0.4 respectively, the blue dot-dashed lines
correspond to unoscillated neutrinos.  A comparison between Figs. 8
and 9 reveals an interesting feature.  The case with the higher value
of $\Delta m^2_{31}$ has a spectral distortion which is less than that
with the lower value, especially when $E$ is between 3 and 5 GeV and 
when $E$ is between 5 and 7 GeV, which are statistically the most
important energy bins.  The reason is that the oscillations are faster for
$\Delta m^2_{31} = 4.0 \times 10^{-3} \ {\rm eV}^2$, leading to
partial averaging and a smaller spectral distortion in the resultant
signal.  In fact, for the same reason, matter effects also are
slightly more important for $\Delta m^2_{31} = 2.0 \times 10^{-3} \
{\rm eV}^2$, as compared to the higher value quoted above.  Such an
increased averaging, especially in the statistically more relevant
lower energy bins, results in a slight fall in the sensitivity of ICAL
to the precision of the oscillation parameters.  This happens 
since the use
of spectral distortion is a crucial component of the latter in
relation to oscillation parameters, 
including the deviation $D$ from the
maximality of $\sin^2\theta_{23}$.

\section{\label{sec:octant}
Sensitivity to the octant of $\theta_{23}$}

We now come to the determination of the sign of $D = 1/2 -
\sin^2\theta_{23}$ which decides whether $\theta_{23} < \pi/4$ ($D$
positive) or $\theta_{23} > \pi/4$ ($D$ negative), i.e. whether
$\theta_{23}$ lies in the first or second octant of the $(0,2\pi]$
range.  In a two-generation picture, the dependence of the survival
probability $P_{\mu\mu}$ only on $\sin^22\theta_{23}$ makes it
impossible to fix the octant of $\theta_{23}$.  Even in a
three-generation framework, given that $\theta_{13}$ and $\alpha$ are
small, the dominant part of $P_{\mu\mu}$ in vacuum is still found to
depend only on $\sin^22\theta_{23}$.  Therefore, once again the 
fixation of the octant of
$\theta_{23}$ would be nearly impossible from a study of
vacuum oscillations in surviving muon neutrinos.  This degeneracy,
however, is broken in matter owing to an additional strong
$\sin^4\theta_{23}$ dependence in $P_{\mu\mu}$, as clear from the
third term $P^3_{\mu\mu}$ in Eq. (\ref{eq:p22term3}).  
Since we have already
demonstrated the significant potentiality of the ICAL-like detector in
deciphering the effects of earth matter on its observed muon
neutrino/antineutrino events, we expect 
such a detector to be sensitive
to the octant of $\theta_{23}$.  That is what we now numerically
investigate.

The simulated event spectrum at this detector 
is first generated at some
nonmaximal ``true'' value of $\theta_{23}$.  Then we try and see if
the wrong octant of $\theta_{23}$ can be ruled out at a good C.L.
Here we 
use anticipated constraints on $\sin^2\theta_{13}$ from future
reactor and long baseline experiments, but we have allowed all other
parameters except $\sin^2\theta_{23}$ to vary freely in the fit.  
We consider three different cases of 
$\sat$ values, namely 0.42, 0.46 and 0.54.
The ``true'' values of the other oscillation parameters are as listed
in Table \ref{tab:benchmark} 
with an assumed normal mass ordering.  
Fig. \ref{fig:delchioctant} shows the results 
of our statistical analysis based on simulated data from 1 MtonY of ICAL 
exposure with the left, middle and right panels corresponding to the values 
0.42, 0.46 and 0.54 respectively of $\sin^2\theta_{23}$(true). 
In each panel the long dashed magenta, short dashed blue, dot-dashed
green, dotted red and solid black lines show the
$\chi^2$ for $\sin^2\theta_{13}$(true) $= 0.00$, 0.01, 0.02, 0.03 
and 0.04 respectively. 
For every nonmaximal
$\sin^2\theta_{23}$(true), there exists a $\sin^2\theta_{23}$(false)
which is given by 
\be
\sin^2\theta_{23} ({\rm false}) = 1 - \sin^2 \theta_{23} ({\rm true})
\ee
on the other side of $\pi/4$.  
For a vanishing $\sin^2\theta_{13}$(true)
there are no matter effects and the $\chi^2$ corresponding to 
both the true and false values
of $\sin^2 \theta_{23}$ are the same. Hence they are 
allowed at the same C.L. and
one fails to fix the octant of $\theta_{23}$ in this case.
However, for
$\scht \neq 0$, matter effects bring in an octant
sensitivity and a false $\sin^2\theta_{23}$ solution can be ruled out,
provided $D$(true) is not too close to zero. 
For a given $\scht$, the C.L. at which this can 
be done for our illustrative cases of $\sat$ 
can be read out using Fig. \ref{fig:delchioctant}, 
from the difference in the $\chi^2$ between 
the true and false solutions.

In order to obtain the limiting value of $\sat$ 
which could still allow for the determination of $sgn(D)$
we define
\be
\Delta \chi^2 &\equiv& \chi^2 (\sin^2\theta_{23} ({\rm true}),
\sin^2\theta_{13} ({\rm true}), {\rm others(true)}) \nonumber \\[2mm] 
&& \hspace*{1cm} - \chi^2(\sin^2\theta_{23} ({\rm
false}),\sin^2\theta_{13}, {\rm others}),
\label{Eq:chioctant}
\ee
with `others' comprising $\Delta m^2_{31}$, $\Delta m^2_{21}$,
$\sin^2\theta_{12}$ and $\delta$.  These, along with
$\sin^2\theta_{13}$, are allowed to vary freely in the fit.  
Eq. (\ref{Eq:chioctant})
gives us a measure of the C.L. at which $\sin^2\theta_{23}$(false) is
disfavored for a given $\sin^2\theta_{23}$(true). 
Fig. \ref{fig:delchioctanttrue} 
shows $\Delta\chi^2$ as a function of 
$\sin^2\theta_{23}$(true) for four
values 0.01, 0.02, 0.03 and 0.04 of $\sin^2\theta_{13}$(true), 
corresponding respectively to the blue dashed, and green dot-dashed,
the red dotted and the black solid lines.  The left hand panel 
shows the results for a normal mass ordering of neutrinos
while the right hand panel corresponds to the situation if the 
mass ordering for the neutrinos were inverted.
The range of
$\sin^2\theta_{23}$(true), for which $\sin^2 \theta_{23}$(false) can
be ruled out at the $3\sigma$ level, is visible from the figure.  In
particular, for a normal neutrino mass ordering,
$\sin^2\theta_{23}$(false) should be excludable at the
$3\sigma$ level from 1 MtonY of ICAL exposure for the following cases:
%
%
%
\be
\sin^2\theta_{23} ({\rm true}) < 0.361 \ {\rm or} \ > 0.633 \ {\rm
for} \ \sin^2\theta_{13} ({\rm true}) = 0.01, \\[2mm]
\sin^2\theta_{23} ({\rm true}) < 0.402 \ {\rm or} \ > 0.592 \ {\rm
for} \ \sin^2\theta_{13} ({\rm true}) = 0.02, \\[2mm]
\sin^2\theta_{23} ({\rm true}) < 0.415 \ {\rm or} \ > 0.580 \ {\rm
for} \ \sin^2\theta_{13} ({\rm true}) = 0.03,\\[2mm]
\sin^2\theta_{23} ({\rm true}) < 0.421 \ {\rm or} \ > 0.573 \ {\rm
for} \ \sin^2\theta_{13} ({\rm true}) = 0.04.
\ee

For an inverted mass ordering, the sensitivity of ICAL
to $sgn(D)$ is seen to be slightly less from 
Fig. \ref{fig:delchioctanttrue}. 
This is the result of the 
following fact. When the mass ordering is inverted,
larger matter effects appear for $\bar\nu_\mu$'s
instead of $\nu_\mu$'s at baseline lengths $L > 1000$ km.
Owing to smaller interaction cross-sections for the 
antineutrinos, we roughly expect only half as many 
$\anumu$-induced antimuon events in the detector 
as compared with
$\numu$-induced 
muon events. Therefore,
the statistical power of the experiment for deciphering $sgn(D)$ 
goes down by roughly a factor of half in this case.

In order to quantify the importance of distinguishing the 
$\numu$-induced muon events from the $\anumu$-induced antimuon 
events by means of the magnetic field in the detector, we have 
repeated our $\chi^2$ 
analysis by ``switching off'' the magnetic field --
or in other words by taking the sum of the muon and antimuon events 
in each bin. Therefore, in this case we have just a 60 bin data. 
Such a non-magnetized iron detector could then rule out the fake 
octant solution at \sig{} level for the normal mass ordering if
$\sin^2\theta_{23} ({\rm true}) < 0.372 (>0.625)$ 
for $\scht=0.04$ 
or $\sin^2\theta_{23} ({\rm true}) < 0.398 (>0.600)$ 
for $\scht=0.04$. It is not surprising that the potential to rule
out the fake octant deteriorates 
significantly in this case, since the matter effects
crucial for the octant sensitivity get diluted if neutrino and 
antineutrino events were added together.
Distinguishing the neutrinos from the antineutrinos is therefore 
extremely important for the physics potental of experiments 
sensitive to large matter effects.

\section{\label{sec:conclusions}Discussions and conclusions}

The measurement of both the magnitude and sign of   
the deviation $D$ of $\sa$ from its maximal value 0.5 is of 
utmost theoretical importance. The best current limit on 
this parameter $D$ comes from the SK atmospheric neutrino 
experiment giving $|D| \leq 0.16$ at the \sig{} level \cite{SKatm}.
More precise measurements of $\theta_{23}$ 
and hence of $D$ are expected from 
future atmospheric neutrino data both from the currently running 
SK experiment and from the planned Megaton 
water Cerenkov detectors. Significantly
better constraints are expected 
from data that will emerge from forthcoming long baseline experiments,
owing to their larger statistics and lower systematic errors.
However, in both these classes of experiments,
$\theta_{23}$ will be mainly determined by the 
$\numu$ (and/or $\anumu$) disappearance channel in vacuum,
which predominantly 
depends on $\sin^22\theta_{23}$. This leads to two very important 
consequences. The first one relates to the fact that the mixing angle 
$\theta_{23}$ is very close to being maximal. The fact that
$\delta(\sa)$ equals $\delta(\sin^22\theta_{23})/(4~\cos2\theta_{23})$
has the following implication. Even though one 
could determine the value of
$\sin^22\theta_{23}$ at the percentage level from the next generation 
long baseline experiments, 
the uncertainty 
in $\sa$ would still remain in the region of 10-20\%, depending 
on $\sat$. The second consequence of the predominant 
$\sin^22\theta_{23}$ dependence of the disappearance probability in 
long baseline experiments means that they are almost insensitive 
to the octant of $\theta_{23}$ and hence to the sign of $D$. 

In this paper we have argued in a quantitative way
that the presence of matter effects in the neutrino survival probability
enhances its sensitivity to $\theta_{23}$. In particular,
we have shown that the strong $\sa$ dependence of the matter effects
in $\pmm$ can be used to increase the sensitivity of $\numu$'s and 
$\anumu$'s disappearance experiments to $D$. Using the signal for 
atmospheric $\numu$ and $\anumu$ in a large magnetized iron 
calorimetric detector like the planned ICAL at INO in India, we 
have demonstrated the followoing fact: for large enough values of
$\sin^2\theta_{13}$(true), the detection of
matter effects in the data would be feasible
and would lead to better constraints on $|D|$.
Appropriate binning of the data in both energy and zenith angle 
holds the key to the observation of matter effects in the 
resultant signal. We have distributed our simulated 
events into twelve zenith angle bins covering both up and down going 
neutrinos
-- each of which was then further divide into five energy bins, giving 
us sixty bins of data. 
The charge discrimination capability of a magnetized
detector like ICAL will allow the separation 
between $\nu_\mu$- and $\bar\nu_\mu$-induced events.
Hence we have simulated separate data sets
for $\numu$'s and $\anumu$'s. Since for a given neutrino mass ordering, 
substantial matter effects appear in only either the neutrino 
or the antineutrino channel, 
the effective use of separate data sets for $\nu_\mu$ and 
$\bar\nu_\mu$ will enable the extraction of matter
effects in the survival probability through a statistical analysis.
We have defined a $\chi^2$ 
function to analyze the data set expected from
atmospheric neutrino events collected in an ICAL-like
detector and presented the
C.L. limits on $|D|$. We have noted that the presence of matter effects 
increase the sensitivity of ICAL to $|D|$ somewhat. For 
$\scht=0.04$, matter effects are seen to improve the \sig{} 
limit on $|D|$ from 0.090 to 0.084. We expect to measure
$|D|$ within 18\% (17\%) at \sig{} with 1 MtonY data when 
$\scht=0.02$ (0.04). This is comparable to the limit on $|D|$ 
expected from the forthcoming long baseline experiments and 
slightly better than what is expected from atmospheric 
neutrino experiments with Megaton water Cerenkov detectors.

Matter effects in $\pmm$ 
open up a new utility for an ICAL-like detector -- sensitivity
to the octant of $\theta_{23}$. As is well known, 
the atmospheric neutrino data set collected so far by the SK 
experiment constrains the atmospheric neutrino 
mixing angle in the form $\sin^22\theta_{23}$ and is hence 
insensitive to the octant of $\theta_{23}$ if its 
true value were non-maximal
This ambiguity in whether $\theta_{23}$ is smaller or greater 
than $\pi/4$ leads to an additional two-fold degeneracy \cite{fltheta23deg}
in the 
measurement of the mixing angle $\theta_{13}$ and the CP phase 
$\delta$ in long 
baseline experiments, looking for electron neutrino appearance in 
an original muon neutrino beam through the $P_{\mu e}$ channel
which depends on $\sa$ and hence to the octant of $\theta_{23}$.
The $\pmm$ channel in the 
long baseline experiments, which is expected to give the best limits 
on $\theta_{23}$, is almost insensitive to its octant 
\cite{hubert2k+hk}. The proposed ways to tackle the 
$\theta_{23}$ octant ambiguity in the 
determination of $\theta_{13}$ and
$\delta$ in long baseline experiments 
include 
(1) combining the data from the $P_{\mu e}$ 
(or $P_{ e \mu }$) channel of the
long baseline experiment with 
data from the next generation reactor experiments 
\cite{react13},
(2) combining the data from the $P_{\mu e}$ 
(or $P_{ e \mu }$) channel of the
long baseline experiments with different 
energies and baseline lengths
and (3) combined studies
of the $P_{\mu e}$ or $P_{ e \mu }$ 
(golden) and the $P_{e\tau}$ (silver) channels \cite{yasuda23}.

The octant ambiguity can be resolved from a direct measurement of the
sign of $D$ utilizing atmospheric neutrino data.
It was shown in \cite{maltonimax23} that the atmospheric neutrino
data in SK like experiments
could be sensitive to the octant of $\theta_{23}$ through the 
$\ms$ driven sub-dominant oscillations. The latter are sizable for 
very small values of the neutrino energies and result in an excess 
in the sub-GeV electron sample in Megaton water Cerenkov detectors. 
They conclude that, with a statistics fifty times the current SK 
statistics, this effect can be used to determine the correct 
octant of $\theta_{23}$ at the \sig{} level if the true value
$\sin^2\theta_{23}{\rm (true)} <0.36$ or $> 0.62$.
We have shown in this paper that for large enough $\scht$,
the observation of matter effects in atmospheric 
neutrinos in an ICAL like detector could be used very effeciently 
to determine the octant of $\theta_{23}$.
The observation of significant matter effects in $\pmm$ 
allows the rejection of the false octant solution of $\theta_{23}$
at the \sig{} level for $\sat < 0.40$ ($< 0.42$)
or $\sat > 0.59$ ($> 0.57$) for $\scht=0.02$ (0.04)
with 1 MtonY statistics in ICAL, given a normal 
neutrino mass ordering. For an inverted mass ordering, 
the false $\theta_{23}$ can be rejected at the 
\sig{} level for $\sat < 0.37$ ($< 0.40$)
or $\sat > 0.62$ ($> 0.60$) for $\scht=0.02$ (0.04). The 
sensitivity to $sgn(D)$ in ICAL is  
therefore slightly less for an inverted mass ordering
than for a normal one. But it is
still better than what is expected from other experiments.
Hence, using matter effects to pick the right octant in an INO-like
experiment seems to be the most promising 
way of resolving the $\theta_{23}$ octant 
ambiguity by determining the sign 
of $D$.

\vskip 1cm
{\small We thank S. R. Dugad, S. Goswami,
M. V. N. Murthy, D. P. Roy and A. Raychaudhuri
for their helpful comments.}






\begin{figure}[t]
\begin{center}
\includegraphics[width=14.0cm, height=8.5cm]
{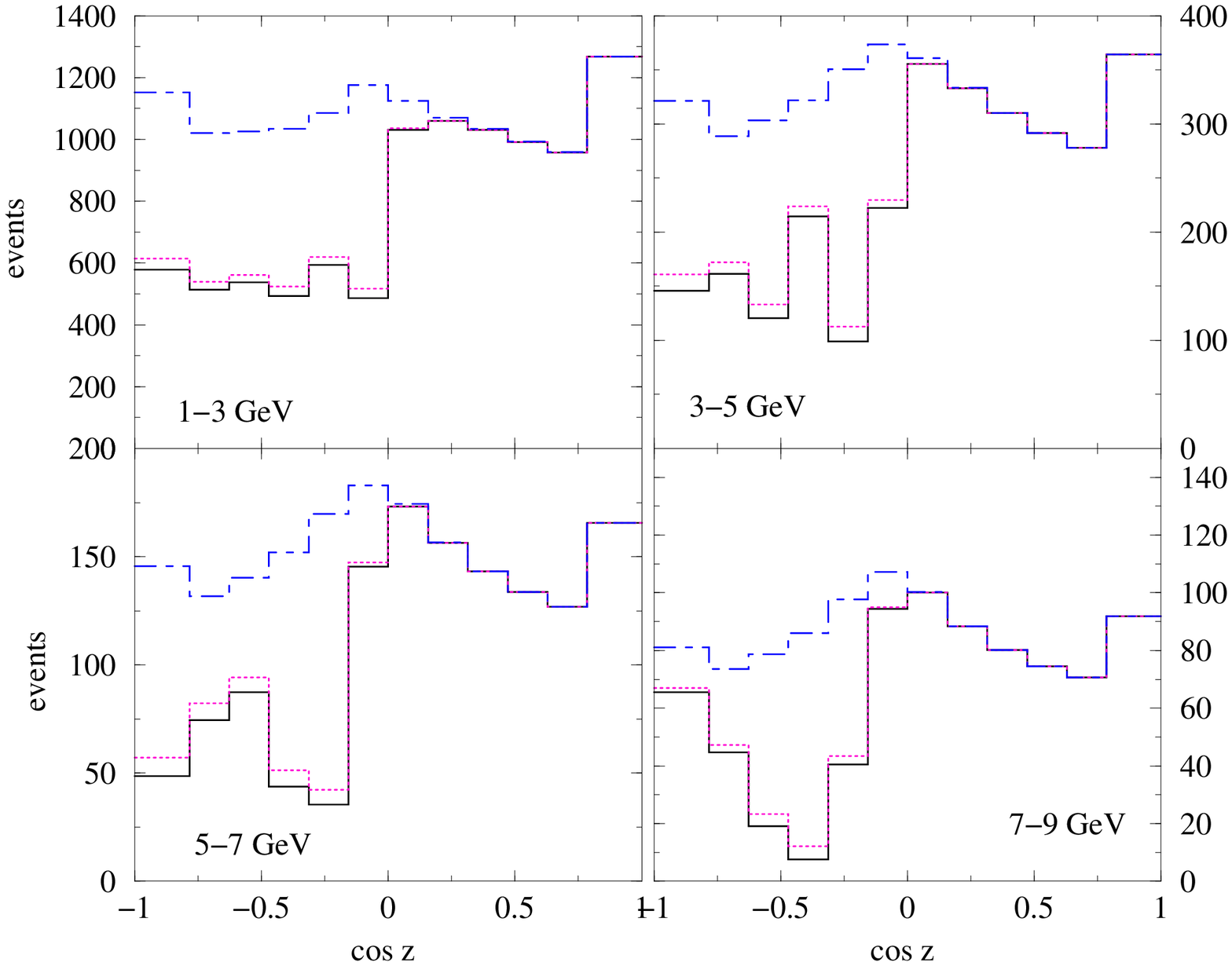}
\caption{
The zenith angle distribution of muon events for four energy 
bins shown in the four panels and for $\ma=2\times 10^{-3}$ eV$^2$.
The black solid lines show the events for maximal $\sa$, the 
dotted magenta lines show the events for $\sa=0.4$ and 
the blue dot-dashed lines show the events for the unoscillated 
flux.
}
\label{fig:ev002}
\end{center}
\begin{center}
\includegraphics[width=14.0cm, height=8.5cm]
{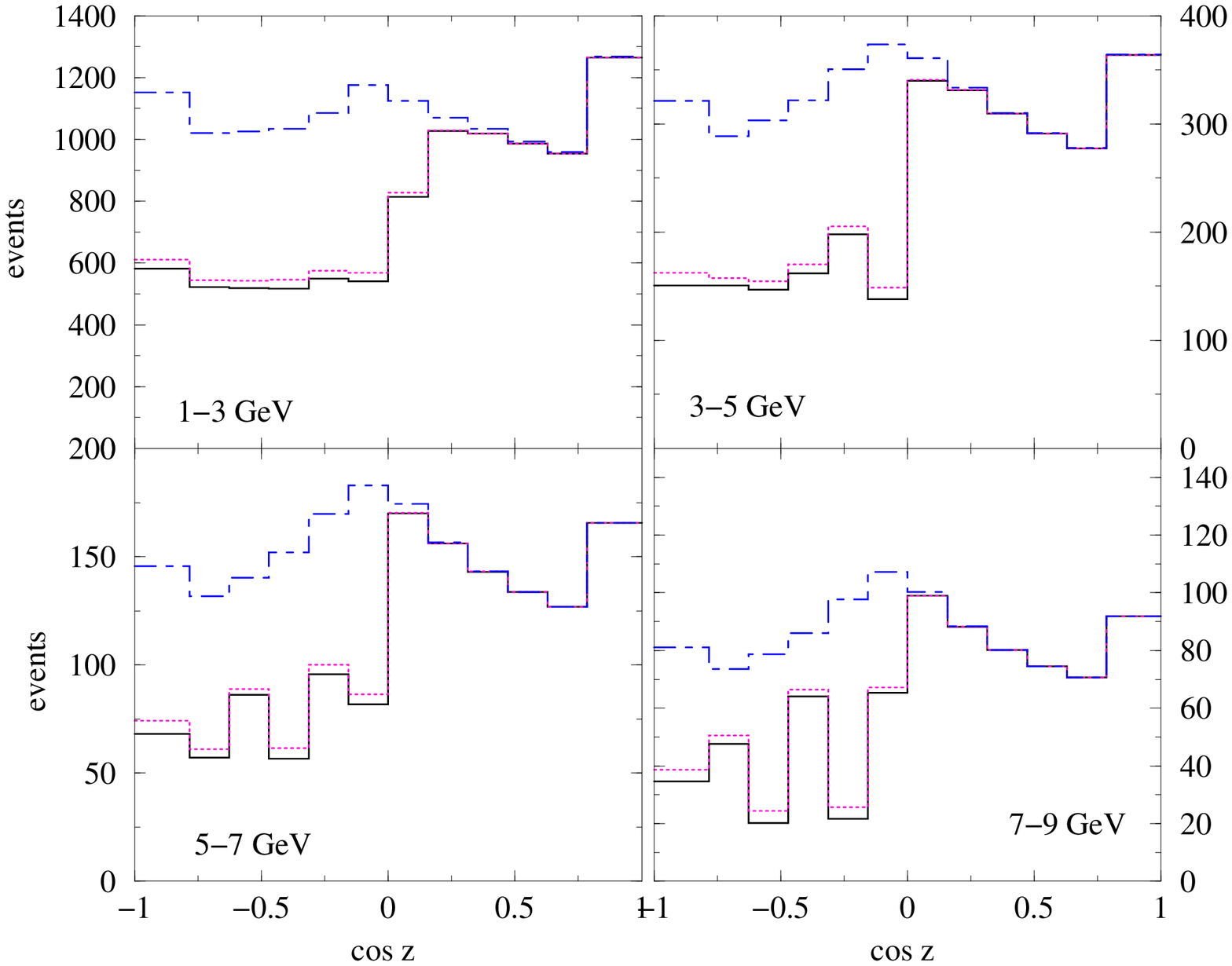}
\caption{
Same as Fig. \ref{fig:ev002}, but for 
$\ma=4\times 10^{-3}$ eV$^2$.
}
\label{fig:ev004}
\end{center}
\end{figure}

\begin{figure}[t]
\begin{center}
\includegraphics[width=16.0cm, height=12.0cm, angle=0]
{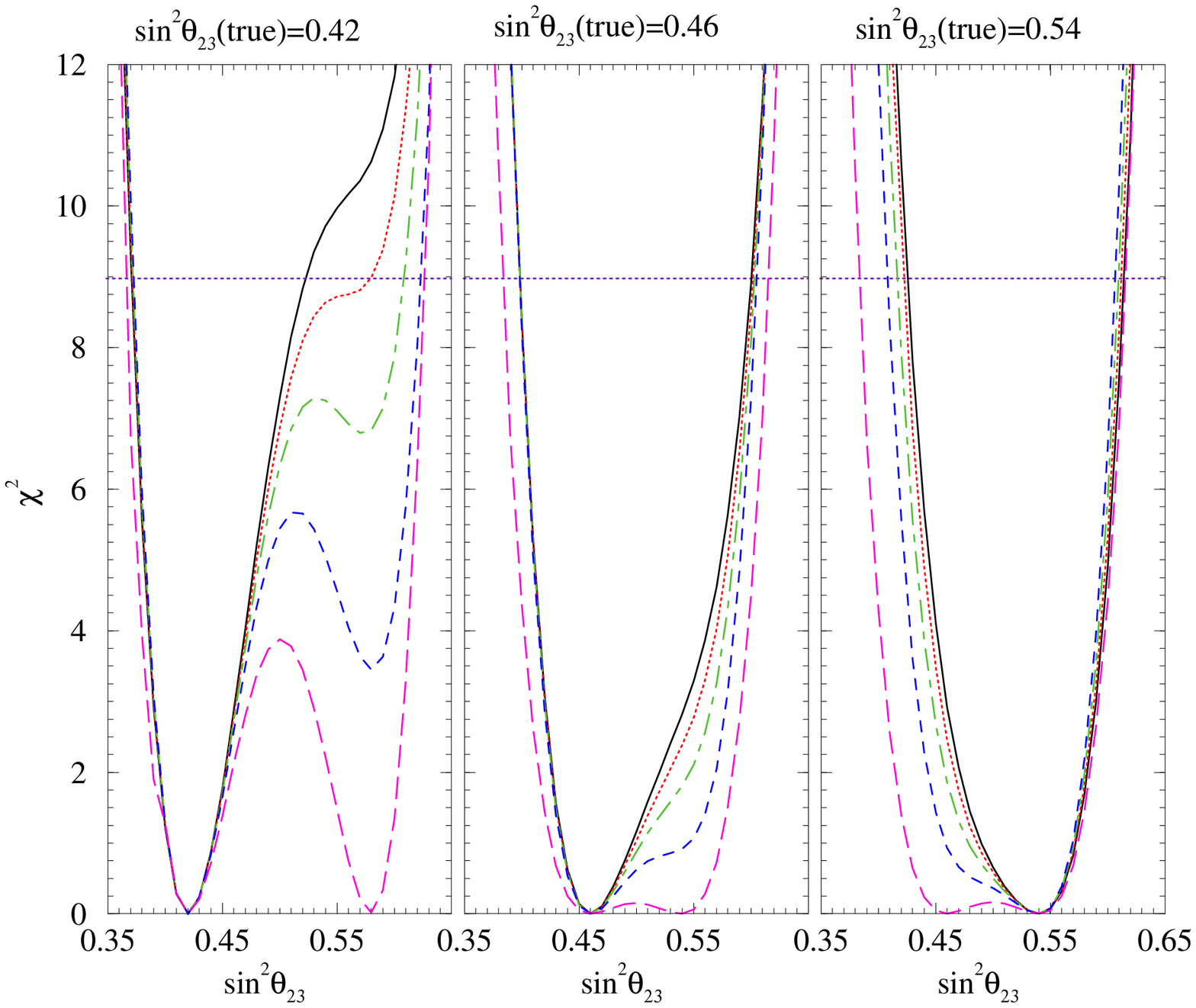}
\caption{
$\chi^2$ as a function of $\sa$, showing the 
``octant sensitivity'' of ICAL. 
The three panels 
project the expected sensitivity of ICAL for 
three different true value of $\sa$: $\sa{\rm (true)}=0.42$ (left panel),
$\sa{\rm (true)}=0.46$ (middle panel) and $\sa{\rm (true)}=0.54$ 
(right panel). The magenta long-dashed lines, blue short-dashed lines,
green dot-dashed lines, red dotted lines and black solid lines
are for $\sch{\rm (true)}=0.00$, 0.01, 0.02, 0.03 and 0.04 respectively.
}
\label{fig:delchioctant}
\end{center}
\end{figure}
\begin{figure}[t]
\begin{center}
\includegraphics[width=16.0cm, height=12.0cm, angle=0]
{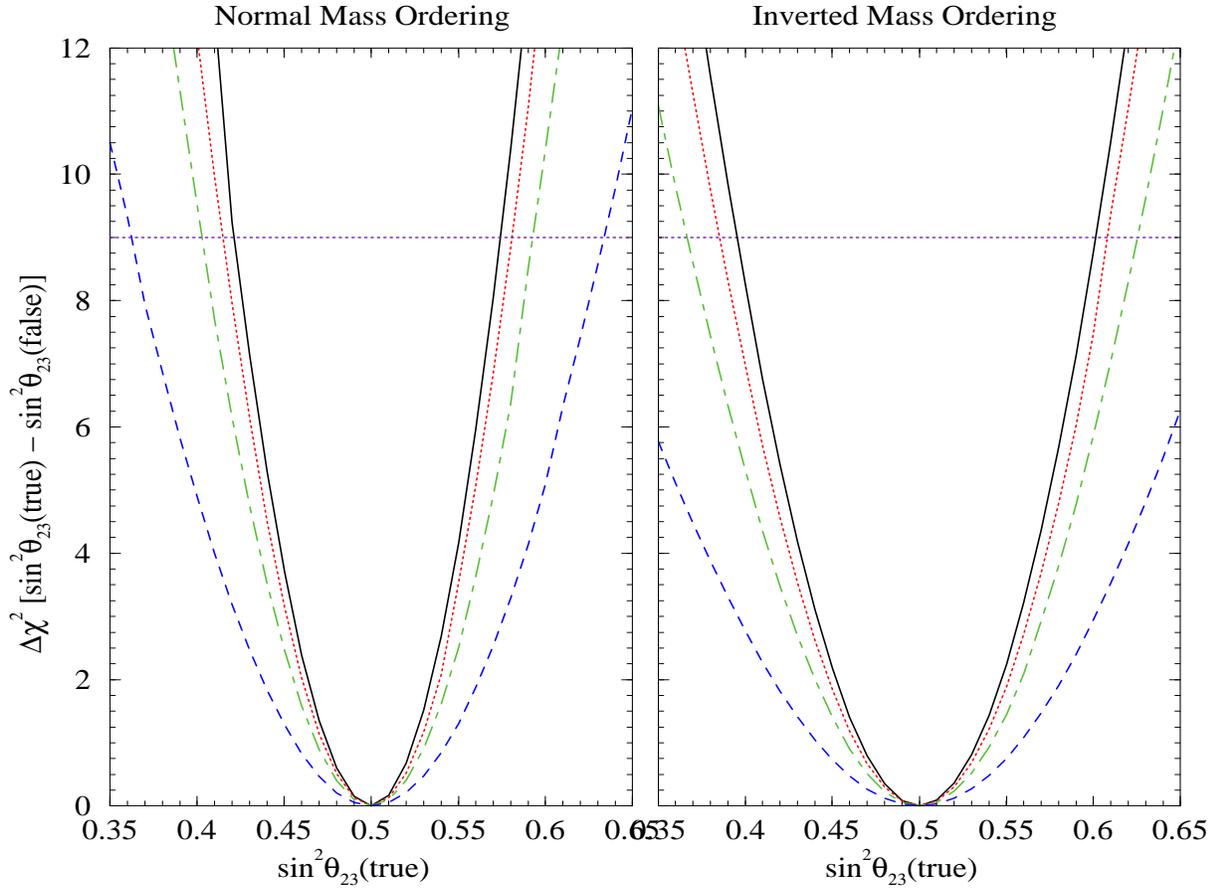}
\caption{
$\Delta \chi^2$ as a function of $\sat$, showing the 
octant sensitivity of ICAL for the normal (left panel) and 
inverted (right panel) neutrino mass ordering.
 The blue short-dashed lines,
green dot-dashed lines, red dotted lines and black solid lines
are for $\sch{\rm (true)}=$ 0.01, 0.02, 0.03 and 0.04 respectively.
}
\label{fig:delchioctanttrue}
\end{center}
\end{figure}

\end{document}